\documentclass[english,11pt]{article}
\usepackage{graphicx}
\usepackage{amsmath,amsbsy,amssymb,amsthm,constants,enumitem,citesort,cite,comment}
\theoremstyle{plain}
\newtheorem{theorem}{\protect\theoremname}
\theoremstyle{plain}
\newtheorem{definition}[theorem]{\protect\definitionname}
\theoremstyle{plain}
\newtheorem{proposition}[theorem]{\protect\propositionname}
\theoremstyle{plain}

\theoremstyle{plain}
\newtheorem{lemma}[theorem]{\protect\lemmaname}

\def \M {\mathbb{M}}
\def \R {\mathbb{R}}
\def \T {\mathbb{T}}
\def \N {\mathbb{N}}

\usepackage{babel}
  \providecommand{\definitionname}{Definition}
  \providecommand{\examplename}{Example}
  \providecommand{\lemmaname}{Lemma}
  \providecommand{\propositionname}{Proposition}
\providecommand{\theoremname}{Theorem}

\title{What Happens to a Manifold Under a Bi-Lipschitz Map?}

\author{Armin Eftekhari\footnote{Institute for Computational Engineering and Sciences (ICES), University of Texas at Austin. Email: armin.eftekhari@utexas.edu} ~and~Michael B. Wakin\footnote{Department of Electrical Engineering and Computer Science, Colorado School of Mines. Email: mwakin@mines.edu \newline \indent This work was partially supported by NSF CAREER Grant CCF-1149225 and NSF Grant CCF-1409258.}}

\begin{document}

\maketitle

\vspace{-0.2in}

\begin{abstract}
We study geometric and topological properties of the image of a smooth submanifold of $\mathbb{R}^{n}$ under a bi-Lipschitz map to $\mathbb{R}^{m}$. In particular, we characterize how the dimension, diameter, volume, and reach of the embedded manifold relate to the original. Our main result establishes a lower bound on the reach of the embedded manifold in the case where $m \le n$ and the bi-Lipschitz map is linear. We discuss implications of this work in signal processing and machine learning, where bi-Lipschitz maps on low-dimensional manifolds have been constructed using randomized linear operators.
\end{abstract}

\noindent {\bf Keywords.} Manifolds, reach, bi-Lipschitz maps, compressive sensing, random projections.

~

\noindent {\bf AMS Subject Classification.} 28A75, 53A07, 57R40, 68P30, 94A12.

\section{Introduction\label{sec:Introduction}}

Let $\mathbb{M}\subset\R^{n}$ be a $k$-dimensional smooth \emph{submanifold} of $\mathbb{R}^{n}$ with finite diameter and volume, and with nonzero \emph{reach}.\footnote{Reach measures the regularity of a manifold. This, as well as more specific requirements on $\M$, are described in Section \ref{sec:Necessary-Concepts}.} Consider also $\Phi_{l,u}:\R^{n} \rightarrow \R^{m}$ that acts as a \emph{bi-Lipschitz} map on $\mathbb{M}$ in the sense that
\begin{equation}
l\cdot\|x-y\|_{2}\le\left||\Phi_{l,u}(x)-\Phi_{l,u}(y)\right\Vert _{2}\le u\cdot\|x-y\|_{2},\qquad\forall x,y\in\M,\label{eq:iso}
\end{equation}
for some $0<l\le1\le u<\infty$. In particular, when $l,u\approx1$, then $\Phi_{l,u}(\cdot)$ is a \emph{near-isometry} on $\M$, in that it barely distorts the pairwise Euclidean distances between points on $\M$.\footnote{The deviation $\max[u-1,1-l]$ (if positive) is commonly referred to as the \emph{isometry constant} of the map $\Phi_{l,u}(\cdot)$ on $\M$.} Such maps naturally arise in a variety of applications in data sciences, often involving dimensionality reduction  \cite{broomhead2001whitney,donoho2006compressed,candes2008restricted,clarkson2008tighter,hegde2008random,eftekhari2015new,hegde2015numax}; we will expand on this remark in Section \ref{sec:Motivation}. Let us point out that we only assume the map $\Phi_{l,u}(\cdot)$ to be bi-Lipschitz on $\M$, not on the entire domain $\R^{n}$. This is a subtle but crucial detail which will be significant, for example, in applications where $\Phi_{l,u}$ is linear and where $m < n$. In such applications $\Phi_{l,u}(\cdot)$ will have a nullspace and map many vectors to $0$, but for~\eqref{eq:iso} to be satisfied with $l > 0$ no vector of the form $x-y$ with $x, y \in \M$, $x \neq y$ can be in the nullspace of $\Phi_{l,u}(\cdot)$.

The setup above naturally raises the following question: What happens to $\M$ under the bi-Lipschitz map $\Phi_{l,u}(\cdot)$? It is in fact trivially verified that $\Phi_{l,u}(\M)\subset\R^{m}$ is itself a smooth submanifold of $\mathbb{R}^{m}$, where $\Phi_{l,u}(\M)$ is the image of $\M$ under $\Phi_{l,u}(\cdot)$. This allows us to define our inquiry more precisely:
\begin{itemize}
\item \textbf{Question:} How do key characteristics of $\Phi_{l,u}(\M)$---namely dimension, diameter, volume, and reach---relate to those of $\M$?
\end{itemize}
When $l,u\approx1$, we will verify that
\begin{align*}
\dim\left(\Phi_{l,u}(\M)\right) & =\dim\left(\M\right), \\
\mbox{diam}\left(\Phi_{l,u}(\M)\right)&\approx\mbox{diam}\left(\M\right), \\
\mbox{vol}_{k}\left(\Phi_{l,u}(\M)\right)&\approx\mbox{vol}_{k}\left(\M\right),
\end{align*}
as formalized in Proposition \ref{lem:easy}. However, for reach---an important attribute of the manifold---the story is somewhat different as additional restrictions must be imposed on $\Phi_{l,u}(\cdot)$ for reach to be approximately preserved. We briefly outline this peculiar case next and defer the details to Section~\ref{sec:Main-Results}.

Reach, as a measure of regularity of a submanifold of Euclidean space, can be traced back to the pioneering works in \emph{geometric measure theory}~\cite{federer1959curvature}. More recently, the concept of reach has proved indispensable in the role that manifold models play in signal processing and machine learning~\cite{niyogi,baraniuk2009random,hegde2008random,yap2013stable,iwen2013approximation,verma2012,davenport2007sfc}. As elaborated later in Proposition \ref{lem:exact iso}, reach is actually preserved under an \emph{exact isometry}, i.e., $\mbox{rch}(\Phi_{1,1}(\M))=\mbox{rch}(\M)$. However, there is generally no relation between $\mbox{rch}(\Phi_{l,u}(\M))$ and $\mbox{rch}(\M)$ when $l<1<u$, and we provide a concrete example in Proposition \ref{fact:In-general,-reach} to demonstrate this. This example involves a nonlinear choice of $\Phi_{l,u}(\cdot)$. Interestingly, however, if we limit ourselves to the class of {\em linear} maps, reach is nearly preserved. Specifically, we can lower bound $\mbox{rch}\left(\Phi_{l,u}(\M)\right)$ in terms of $\mbox{rch}(\M)$ when $\Phi_{l,u}(\cdot)$ is a linear near-isometry; we formalize this claim in Theorem~\ref{thm:(Near-Isometry) rch preserved}.

Before attending to the details, however, we present the necessary background in Section~\ref{sec:Necessary-Concepts} to make this paper self-contained. Then, Section~\ref{sec:Main-Results} makes precise the claims we outlined above, with further discussion and context for these results appearing in Section~\ref{sec:discussion}. In Section~\ref{sec:Motivation}, revisiting  the applications of nearly-isometric maps in data sciences, we show how the (seemingly abstract) Question above naturally arises in modern signal processing and machine learning, and we discuss the implications of our results in these contexts. We defer all proofs to the appendices.

\section{Necessary Concepts\label{sec:Necessary-Concepts}}

Let $\M\subset\mathbb{R}^{n}$ be a smooth, bounded, and boundary-less $k$-dimensional submanifold of $\mathbb{R}^{n}$. More formally, $\mathbb{M}$ satisfies the following. Any point $x\in\mathbb{M}$ belongs to some \emph{neighborhood} $\mathbb{U}_{x}\subseteq\mathbb{M}$ that is \emph{diffeomorphic}\footnote{A diffeomorphism is a smooth and bijective map with a smooth inverse.} to $\mathbb{R}^{k}$, and the transition between adjacent neighborhoods is smooth \cite{morgan1998riemannian,lee2009manifolds}.\footnote{By smooth transition we mean the following: If $\phi_{x}:\mathbb{U}_{x}\rightarrow\mathbb{R}^{k}$ and $\phi_{x'}:\mathbb{U}_{x'}\rightarrow\mathbb{R}^{k}$ are, respectively, the diffeomorphisms corresponding to the neighborhoods $\mathbb{U}_{x}$ and $\mathbb{U}_{x'}$ on the manifold $\mathbb{M}$ and $\mathbb{U}_{x}\cap\mathbb{U}_{x'}\ne\emptyset$, then the composition $(\phi_{x}\circ\phi_{x'}^{-1})(\cdot)$ must be smooth (wherever defined).}

\begin{figure}[t]
\begin{center}
\includegraphics[width=50mm]{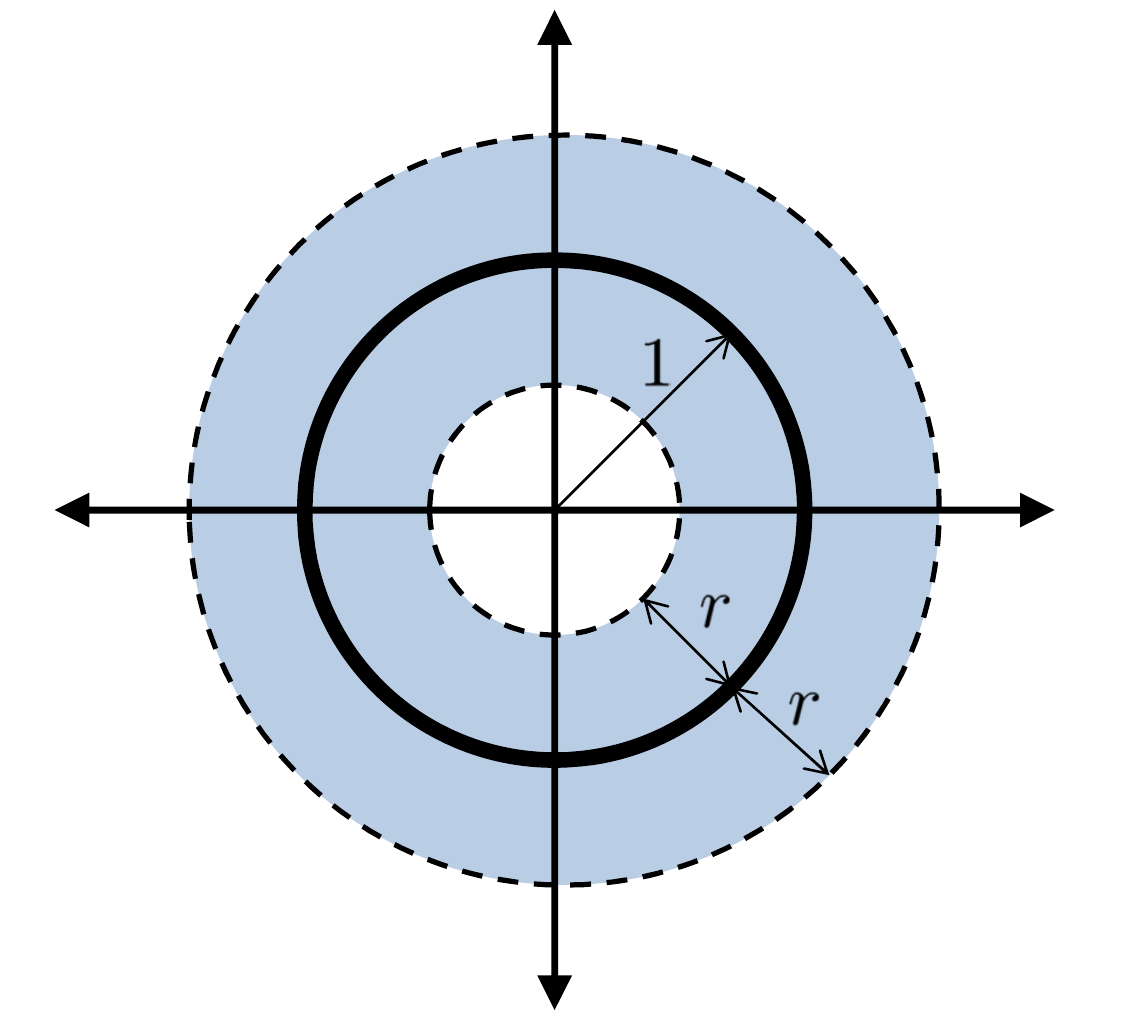}
\end{center}
\vspace*{-5mm} \caption{\small\sl \label{fig:circle} The reach of the unit circle (in bold) is equal to $1$, as the open normal bundle of radius $r$ (shaded) is embedded in $\mathbb{R}^{2}$ for any $r<1$. Moreover, any point in $\R^{2}$ whose distance from the unit circle is less than $1$ has a unique nearest point on the unit circle.
}
\end{figure}

To every point $x\in\mathbb{M}$, we  assign a \emph{tangent subspace} $\mathbb{T}_{x}\mathbb{M}\subset\mathbb{R}^{n}$ comprised of the directions of all smooth curves on $\mathbb{M}$ that pass through $x$:
\begin{align*}
&\mathbb{T}_{x}\mathbb{M}:= \\
&~~~\mbox{span}\left[\left\{ \frac{d\gamma}{dt}(0)\,|\,\gamma:[-1,1]\rightarrow\mathbb{M}\mbox{ is a smooth curve and }\gamma(0)=x\in\mathbb{M}\right\} \right].
\end{align*}
At any $x\in\mathbb{M}$, we note that $\mathbb{T}_{x}\mathbb{M}$ is a $k$-dimensional linear subspace of $\mathbb{R}^{n}$. The \emph{tangent bundle} of $\mathbb{M}$ is the collection of all tangent subspaces:
\[
\mathbb{T}\mathbb{M}:=\bigcup_{x\in\mathbb{M}}\{x\}\times\mathbb{T}_{x}\mathbb{M}.
\]
The \emph{normal subspace} $\mathbb{N}_{x}\mathbb{M}$ is the orthogonal complement of $\mathbb{T}_{x}\mathbb{M}$ with respect to $\mathbb{R}^{n}$ so that
\[
\mathbb{R}^{n}=\mathbb{T}_{x}\mathbb{M}\oplus\mathbb{N}_{x}\mathbb{M},
\]
(with $\oplus$ standing for direct sum). Thus, at any $x\in\mathbb{M}$, $\mathbb{N}_{x}\mathbb{M}$ is an $(n-k)$-dimensional linear subspace of $\mathbb{R}^{n}$. The \emph{normal bundle} of $\mathbb{M}$ is the collection of all normal subspaces:
\[
\mathbb{N}\mathbb{M}:=\bigcup_{x\in\mathbb{M}}\{x\}\times\mathbb{N}_{x}\mathbb{M}.
\]
For $r>0$, we also let $\N^r\M\subset\N\M$ denote the \emph{open normal bundle} of $\M$ of radius $r$ comprised of all normal vectors of length less than $r$. For example, when $\M$ is the unit circle in $\mathbb{R}^{2}$ and $r\in(0,1)$, $\N^r\M$ may be identified with a disc of width $2r$ (around the circle); see Figure~\ref{fig:circle}.

The reach of $\mathbb{M}$ \cite{federer1959curvature}, denoted by $\mbox{rch}(\mathbb{M})$ throughout, is a geometric attribute that captures valuable local and global properties of a manifold (that we will highlight shortly).

\begin{definition}\label{def:rch}
\textbf{{(Reach)}}  For a smooth submanifold $\mathbb{M}\subset\mathbb{R}^{n}$, the reach of $\mathbb{M}$ (denoted by $\mbox{rch}(\mathbb{M})$) is the
largest number having the following property: the open normal bundle of $\mathbb{M}$ of radius $r$ is embedded in $\mathbb{R}^{n}$ for all $r<\mbox{rch}(\M)$.
\end{definition}

For example, the reach of a circle of radius $\rho$ is simply $\rho$ itself. This fact is generalized by the following result, which may be of its own independent interest.

\begin{proposition}
\label{lem:ellipsoidreach}\textbf{\emph{(Reach of an ellipsoid)}}
Let $\M$ be an $(n-1)$-dimensional ellipsoid in $\mathbb{R}^n$ with principal axes $r_1 \ge r_2 \ge \cdots \ge r_n > 0$. Then
\[
 \mbox{rch}(\mathbb{M}) = \frac{r_n^2}{r_1}.
\]
\end{proposition}

By definition, any point in $\mathbb{R}^{n}$ within the distance $\mbox{rch}(\M)$ of $\M$ has a unique nearest point on $\M$. The role of reach can be summarized in two key properties (see the Toolbox in \cite{eftekhari2015new} for more details). The first property concerns local regularity of $\M$: the curvature of any unit-speed {\em geodesic curve}\footnote{Loosely speaking, a geodesic curve between a pair of points on $\M$ is a curve on $\M$ that locally minimizes the distances \cite{lee2009manifolds}. For example, an arc on a circle is a geodesic curve between its end points. Note that the geodesic curve between two points is not necessarily unique.} on $\M$ is bounded by $(\mbox{rch}(\M))^{-1}$. \label{p:curv rch} The second property concerns global regularity: at long \emph{geodesic distances},\footnote{The geodesic distance between a pair of points on $\M$ is the length of the shortest geodesic curve connecting those two points \cite{lee2009manifolds}.} the reach controls how close the manifold may curve back upon itself. For example, supposing $x,y\in\M$ with geodesic distance $d_{\M}(x,y)$, we have that
\[
d_{\M}(x,y)>\mbox{rch}(\M)\Longrightarrow\|x-y\|_{2}>\frac{\mbox{rch}(\M)}{2}.
\]

\section{Results\label{sec:Main-Results}}

Let $\M\subset\R^{n}$ be a smooth, bounded, and boundary-less $k$-dimensional submanifold of $\R^{n}$. We let $\mbox{diam}(\M)$, $\mbox{vol}_{k}(\M)$, and $\mbox{rch}(\M)>0$ denote, respectively, the diameter, $k$-dimensional volume, and reach of $\M$. For $0<l\le1\le u<\infty$ and for $m \ge k$, let $\Phi_{l,u}:\R^{n}\rightarrow\R^{m}$ be a smooth bi-Lipschitz map on $\M$ so that
\begin{equation}
l\cdot\|x-y\|_{2}\le\|\Phi_{l,u}(x)-\Phi_{l,u}(y)\|_{2}\le u\cdot\|x-y\|_{2},\qquad\forall x,y\in\M.\label{eq:near iso}
\end{equation}
Some basic properties of $\Phi_{l,u}(\M)$, the image of $\M$ under $\Phi_{l,u}(\cdot)$, are collected below and proved in Appendix~\ref{sec:Proof-of-Lemma easy}. One key insight used in the proof is the following: Under $\Phi_{l,u}(\cdot)$, a tangent vector $v\in\T_x\M$ is mapped to a tangent vector at  $\Phi_{l,u}(x)$. To be precise, $v\rightarrow D\Phi_{l,u}(x) \cdot v$, where $D\Phi_{l,u}(x)\in\mathbb{R}^{m\times n}$ is the derivative of $\Phi_{l,u}(\cdot)$ at $x$. Moreover, $ D\Phi_{l,u}(x)$ acts as a near-isometry on $\T_{x}\M$ if $\Phi_{l,u}(\cdot)$ is itself a near-isometry on $\M$.

\begin{proposition}
\label{lem:easy}\textbf{\emph{(Basic properties)}} Let the manifold $\M$ and map $\Phi_{l,u}(\cdot)$ be as specified above. Then, $\Phi_{l,u}(\M)$ is itself a smooth, bounded, and boundary-less $k$-dimensional submanifold of $\R^{m}$. Moreover, it holds that
\[
l\cdot \mbox{diam}(\M) \le \mbox{diam}\left( \Phi_{l,u}(\M) \right) \le u\cdot \mbox{diam}(\M),
\]
\[
l^{k}\cdot\mbox{vol}_{k}(\M)\le\mbox{vol}_{k}\left(\Phi_{l,u}\left(\M\right)\right)\le u^{k}\cdot\mbox{vol}_{k}(\M).
\]
\end{proposition}
Further discussion of this result appears in Section \ref{sec:discussion}.

Before we study how a nearly isometric map affects the reach of a smooth manifold, some results are in order to build intuition and form  insight to the problem. The next result, proved in Appendix~\ref{sec:Proof-of-Lemma key exact iso}, states that (not surprisingly) the tangent structure of the manifold remains entirely intact under an exact isometry on $\M$, i.e., when $l=u=1$. The key insight used in the proof is that the angle between any pair of tangent vectors in $\M$  is  preserved in $\Phi_{l,u}(\M)$ if $\Phi_{l,u}(\cdot)$ is an exact isometry on $\M$.

\begin{proposition}
\label{lem:key exact iso}\textbf{\emph{(Tangent bundle under exact isometry)}}
For the manifold $\M$ specified earlier, take $p,q\in\M$. When $l=u=1$, it holds that
\[
\angle\left[\T_{\Phi_{l,u}(p)}\Phi_{l,u}\left(\M\right),\T_{\Phi_{l,u}(q)}\Phi_{l,u}\left(\M\right)\right]=\angle\left[\T_{p}\M,\T_{q}\M\right],
\]
with $\angle[\mathbb{A},\mathbb{B}]$ standing for the angle between the subspaces $\mathbb{A}$ and $\mathbb{B}$.\footnote{\label{p:angle}Angle between subspaces generalizes the notion of angle between lines. The (largest principle) angle between the two subspaces $\mathbb{A}$ and $\mathbb{B}$ is defined such that $\cos(\angle[\mathbb{A},\mathbb{B}]):= \min_a\max_b \cos(\angle [a,b])$, where  the optimization is over all vectors $a\in\mathbb{A}$ and $b\in\mathbb{B}$.
See \cite{horn1994topics} for more details.}
\end{proposition}

To extend Proposition \ref{lem:key exact iso} to a general bi-Lipschitz map on $\M$ with $l\le1\le h$, we restrict ourselves to linear maps. The next result is proved in Appendix \ref{sec:Proof-of-Lemma key near iso}.
\begin{proposition}
\label{lem:key near iso}\textbf{\emph{(Tangent bundle under linear near-isometry)}}
With the manifold $\M$ as before, take $p,q\in\M$. If the map $\Phi_{l,u}:\R^{n}\rightarrow\R^{m}$ is linear on $\R^{n}$ and bi-Lipschitz on $\M$ (i.e., satisfies~(\ref{eq:near iso})), then
\begin{align*}
&\left|\cos\left(\angle[\T_{\Phi_{l,u}(p)}\Phi_{l,u}\left(\M\right),\T_{\Phi_{l,u}\left(q\right)}\Phi_{l,u}\left(\M\right)]\right)-\cos\left(\angle\left[\T_{p}\M,\T_{q}\M\right]\right)\right|
\\
& \qquad \le \frac{25}{4l^2} \cdot \frac{\|p-q\|_2^2}{  \mbox{rch}(\mathbb{M})^2}  \cdot \left(\max(\|\Phi_{l,u}\|^2,u^2)+1 \right)^2 +  \frac{18\max\left(\Delta_{l,u},\sqrt{\Delta_{l,u}}\right)}{l^2},
\end{align*}
where $\Delta_{l,u}:=\max\left[1-l^{2},u^{2}-1\right]$ and $\|\Phi_{l,u}\|$ is the spectral norm of the matrix representation of the linear operator $\Phi_{l,u}(\cdot)$.
\end{proposition}
In words, the angles between tangent spaces on the manifold remain \emph{nearly} unchanged under a linear near-isometry ($l,u\approx1$), for sufficiently close points on the manifold (those within a distance of approximately ${\|\Phi_{l,u}\|^{-2} \mbox{rch}(\mathbb{M})}$). The proof of this fact involves sampling two additional points from the manifold, one near $p$ and one near $q$, controlling the distances between all four of these points under the bi-Lipschitz map $\Phi_{l,u}$, and relating these distances to the angles between the tangent spaces at $p$ and $q$.

Finally, what can be said about $\mbox{rch}(\Phi_{l,u}(\M))$? The next result, proved in Appendix \ref{sec:proof of exact iso}, deals with the special case of an exact isometry.

\begin{proposition}
\label{lem:exact iso}\textbf{\emph{(Reach under exact isometry)}}
Consider the manifold $\M$ as before, and take $l=u=1$. Then,
\[
\mbox{rch}\left(\Phi_{l,u}\left(\M\right)\right)=\mbox{rch}\left(\M\right).
\]
\end{proposition}
We note that Proposition~\ref{lem:exact iso} does not require $\Phi_{1,1}:\R^{n} \rightarrow \R^{m}$ to be linear, nor does it place any restriction on $m$ (aside from the requirement---assumed throughout this section---that $m \ge k$). The key insight here is that $\Phi_{1,1}(\cdot)$ preserves not only pairwise distances on $\M$, but also the angles between pairs of tangent spaces of $\M$ (see Proposition~\ref{lem:key exact iso}). This in turn guarantees that $\Phi_{1,1}(\M)$ is as regular as $\M$ itself (as recorded in the above proposition).

What about the case $l<1$ or $u>1$, where $\Phi_{l,u}(\cdot)$ is not an exact isometry on $\M$? In full generality, there is no relation between $\mbox{rch}(\Phi_{l,u}(\M))$ and $\mbox{rch}(\M)$. This is demonstrated in Appendix~\ref{sec:Proof-of-Proposition in general} where, by adding a smooth ``cusp'' of width $2\delta$ to a line segment in $\R^2$, we explicitly construct a manifold $\M$ and a near-isometric map $\Phi_{l,u}(\cdot)$ on $\M$ such that $\mbox{rch}(\Phi_{l,u}(\M))\le \frac{\delta}{\sqrt{2}}\ll\infty=\mbox{rch}(\M)$.

\begin{proposition}
\textbf{\label{fact:In-general,-reach}\emph{(Negative result)}}
In general, reach is not preserved under a bi-Lipschitz mapping.
\end{proposition}

When $\Phi_{l,u}(\cdot)$ is linear, however, we can lower bound $\mbox{rch}\left(\Phi_{l,u}(\M)\right)$ in terms of $\mbox{rch}(\M)$; the following theorem is proved in Appendix~\ref{sec:Proof-of-Theorem linear Phi}.

\begin{theorem}
\textbf{\label{thm:(Near-Isometry) rch preserved}\emph{(Reach under linear near-isometry)}}
Consider the manifold $\M$ specified earlier. Suppose $m \le n$, and suppose $\Phi_{l,u}(\cdot)$ to be a rank-$m$ linear map from $\R^{n}$ to $\R^{m}$ with its $m$ nonzero singular values in the interval $[\sigma_{\min},\sigma_{\max}]\subset (0,\infty)$. Then, if $m=n$, it holds that
\[
\mbox{rch}\left(\Phi_{l,u}(\M)\right) \ge \frac{\sigma_{\min}^2}{\sigma_{\max}}\cdot\mbox{rch}(\M),
\]
whereas, if $m<n$, it holds that
\[
\mbox{rch}\left(\Phi_{l,u}(\M)\right) \ge \frac{\sigma_{\min}^2l^2}{\sigma_{\max}^3}\cdot\mbox{rch}(\M).
\]
\end{theorem}

We note that Theorem~\ref{thm:(Near-Isometry) rch preserved} does require $\Phi_{l,u}:\R^{n} \rightarrow \R^{m}$ to be linear, and it also requires that $m \le n$ (in addition to the requirement that $m \ge k$). The proof, without any loss of generality, models the linear map $\Phi_{l,u}(\cdot)$ as an orthogonal projection followed by stretching in $\R^m$ along different coordinates, and then records how the reach changes in each of these two steps. In the first step, we use Definition \ref{def:rch} to directly quantify the rather involved effect of orthogonal projection on the reach. On the other hand, qualitatively speaking, modestly scaling an object along different coordinates will not substantially distort its geometry, and we will solidify this notion in the second step of the proof. Combining the two steps proves Theorem \ref{thm:(Near-Isometry) rch preserved}.

\section{Discussion}
\label{sec:discussion}

Manifold embeddings have a rich history in differential topology and geometry. The following discussion might help put our results in the proper context. First, the (strong) Whitney embedding theorem~\cite{whitney1936differentiable} states that any $k$-dimensional smooth manifold $\M$ (not necessarily originating as a submanifold of any Euclidean space) can be smoothly embedded into $\R^{2k}$.\footnote{In this context, a smooth embedding is simply a diffeomorphism.} In contrast, the type of bi-Lipschitz map $\Phi_{l,u}:\R^n\rightarrow\R^m$ that we consider in this paper can be interpreted as \emph{stably} embedding $\M$ (which does originate as a submanifold of $\R^n$) into $\R^m$. In particular, stability in this sense refers to the preservation of extrinsic, Euclidean distances as formalized in~\eqref{eq:iso}. As noted in~\cite{baraniuk2009random}, this type of embedding can also imply the preservation of intrinsic, geodesic distances. The celebrated Nash embedding theorem~\cite{nash1956imbedding} also concerns the embedding of manifolds in Euclidean space, but with an exact isometry. However, isometry in this context refers only to the preservation of intrinsic, geodesic distances.

In Section~\ref{sec:Motivation}, we discuss how certain randomized constructions have been used to construct linear mappings $\Phi_{l,u}:\R^n\rightarrow\R^m$ that satisfy~\eqref{eq:iso} with $m$ as small as possible. Deterministic linear constructions, adapted to the structure of the manifold $\M$ have also been proposed~\cite{broomhead2001whitney,hegde2015numax} to satisfy~\eqref{eq:iso}. Other manifold learning algorithms such as ISOMAP~\cite{tenenbaum2000global}, however, typically involve nonlinear maps designed to preserve only intrinsic, geodesic distances.

Regarding the implications of bi-Lipschitz mappings that we have derived in this paper, let us begin by discussing the assumptions made in Proposition~\ref{lem:easy}. First, it is worth noting that the dimension of a manifold is actually invariant under any homeomorphism.\footnote{A homeomorphism is a continuous and bijective map with a continuous inverse.} One might also consider replacing the bi-Lipschitz map in Proposition \ref{lem:easy} with a ``locally'' bi-Lipschitz map, one that barely distorts the angle between any pair of tangent vectors anchored at the same point on $\mathbb{M}$. It is not difficult to verify that the dimension of $\mathbb{M}$ remains invariant under a locally bi-Lipschitz map and that such a map  only modestly changes the volume. The diameter and reach, however, might change drastically under a locally bi-Lipschitz map. One example is rolling a thin and very long strip of paper along its length (as one would roll up a carpet); this transformation does not distort the angle between any pair of tangent vectors (at the same point) but can arbitrarily reduce the diameter and reach. On the other hand, it is possible to construct an operator that maps such a manifold to itself while significantly distorting the tangent vectors (one could envision placing one's palm on a rubber sheet, and twisting). In this case, the diameter, volume, and reach would all remain the same. However, depending on the conditioning of the Jacobian at a point on the sheet, a pair of tangent vectors at that point could map to vectors with a significantly different angle between them.

Regarding Proposition \ref{lem:key near iso}, we believe that there may be some room for improvement. While this result confirms that the angles between tangent spaces are better preserved as the isometry of the bi-Lipschitz map tightens (i.e., as $\Delta_{l,u} \rightarrow 0$), for a fixed $p$ and $q$, the right hand side of the bound does not go to zero. We do believe that some dependence on $\|p-q\|_2^2$ is natural: for points farther away, the isometry constant must be tighter in order to preserve the angles between the respective tangent spaces. However, it may be possible to derive a bound on the right hand side that scales with the product of $\|p-q\|_2^2$ and $\Delta_{l,u}$, rather than their sum.

Finally, let us explore the tightness of the bounds in Theorem~\ref{thm:(Near-Isometry) rch preserved} using some toy examples. For instance, consider $\M$ to be the unit circle in $\R^2$, and let $\Phi_{l,u}:\mathbb{R}^2\rightarrow\mathbb{R}^2$ be the linear map specified by the matrix
$$
\Phi_{l,u} = \left[
\begin{array}{cc}
3 & 0\\
 0 & \frac{1}{2}
\end{array}
\right].
$$
Then $\sigma_{\min} = 1/2$, $\sigma_{max} = 3$, and $l=1/2$, $u=3$. Here, $\Phi_{l,u}(\M)$ is an ellipse with principal axes $r_1 = 3$ and $r_2 = 1/2$. Thus, by Proposition \ref{lem:ellipsoidreach}, $\mbox{rch}(\Phi_{l,u}(\M))={r_2^2}/{r_1} =1/12$, which precisely matches the lower bound in Theorem \ref{thm:(Near-Isometry) rch preserved}.

As another example, suppose that $\M$ is the unit circle along the plane in $\R^3$ that passes through $e_1=[1,0,0]^T$ and makes an angle $\theta$ with $e_2=[0,1,0]^T$. Let $\Phi_{l,u}:\mathbb{R}^3\rightarrow\mathbb{R}^2$ be the linear map specified by the matrix
$$
\Phi_{l,u} = \left[
\begin{array}{ccc}
1 & 0 & 0\\
0 & 1 & 0
\end{array}
\right].
$$
Then, one can verify that $\sigma_{\min}=\sigma_{\max}=1$, and that $l=\cos\theta$, $u=1$. In this case, $\Phi_{l,u}(\M)$ is an ellipse with principal axes $r_1=1$ and $r_2=\cos \theta=l$. The reach of this ellipse is given by Proposition \ref{lem:ellipsoidreach}:  $\mbox{rch}(\Phi_{l,u}(\M))={r_2^2}/{r_1}=l^2$, which exactly matches the lower bound in Theorem \ref{thm:(Near-Isometry) rch preserved}.\footnote{Note the lower bounds in Theorem \ref{thm:(Near-Isometry) rch preserved} scale linearly if $\Phi_{l,u}(\cdot)$ were to be scaled. As a result, the same conclusion holds true if we replace the map $\Phi_{l,u}(\cdot)$  in this example with, say, $2\Phi_{l,u}(\cdot)$. }

However, it is not difficult to construct an example where the bound in Theorem \ref{thm:(Near-Isometry) rch preserved} is \emph{not} tight. Again suppose that $\M$ is the unit circle along the plane in $\R^3$ that passes through $e_1=[1,0,0]^T$ and makes an angle $\theta$ with $e_2=[0,1,0]^T$. Now, let $\Phi_{l,u}:\mathbb{R}^3\rightarrow\mathbb{R}^2$ be the linear map specified by the matrix
$$
\Phi_{l,u} = \left[
\begin{array}{ccc}
2 & 0 & 0\\
0 & 1 & 0
\end{array}
\right].
$$
Then, one can verify that $\sigma_{\min}=1$, $\sigma_{\max}=2$, and that $l=\cos\theta$, $u=2$. In this case, $\Phi_{l,u}(\M)$ is an ellipse with principal axes $r_1=2$ and $r_2=\cos \theta=l$. The reach of this ellipse is given by Proposition \ref{lem:ellipsoidreach}:  $\mbox{rch}(\Phi_{l,u}(\M))={r_2^2}/{r_1}=\cos^2\theta/2$, which is larger than what Theorem \ref{thm:(Near-Isometry) rch preserved} predicts, namely
\begin{equation*}
\frac{\sigma_{\min}^2l^2}{\sigma_{\max}^3}= \frac{\cos^2\theta}{8}.
\end{equation*}
This appears to be an artifact of the proof technique which, when $m<n$,  decomposes $\Phi_{l,u}(\cdot)$ into two bi-Lipschitz maps with isometry constants of $\frac{l}{\sigma_{\max}}\le \frac{u}{\sigma_{\min}}$ and $\sigma_{\min}\le \sigma_{\max}$, respectively.
%
%More generally, when $\sigma_{\min}<1$ or $\sigma_{\max}>1$, it is very likely that the lower bound in Theorem \ref{thm:(Near-Isometry) rch preserved} is not sharp. This appears to be an artifact of the proof technique which decomposes $\Phi_{l,u}(\cdot)$ into two bi-Lipschitz maps with isometry constants of $\frac{l}{\sigma_{\max}}\le \frac{u}{\sigma_{\min}}$ and $\sigma_{\min}\le \sigma_{\max}$, respectively

\section{Applications in Dimensionality Reduction\label{sec:Motivation}}

We conclude by noting that the Question posed in Section~\ref{sec:Introduction} is strongly motivated by recent advances in signal processing and machine learning. The Information Age has carried with it the burden of acquiring, storing, processing, and communicating increasingly higher dimensional signals and data sets. Fortunately, in many cases, the information contained within a high-dimensional signal or data set actually obeys some sort of concise, low-dimensional model~\cite{cande2008introduction,eftekhari2015new}. Of particular interest to us here is the common scenario where the data lives on a $k$-dimensional submanifold $\M\subset\mathbb{R}^{n}$ (and typically $k\ll n$)~\cite{clarkson2008tighter,eftekhari2015new}.

The low dimension of the manifold model motivates the use of \emph{compressive measurements} for simplifying the data acquisition process. Rather than designing a (possibly very expensive) sensor to measure a signal $x\in\mathbb{R}^{n}$, for example, it often suffices to design a sensor that can measure a much shorter vector $\chi=\Phi_{l,u}x\in\mathbb{R}^{m}$, where $\Phi_{l,u}\in\mathbb{R}^{m\times n}$ is a linear measurement operator that acts as a near-isometry on $\M$ (see (\ref{eq:iso})), and where typically $m \ll n$. Data inference tasks (e.g., classification or parameter estimation) can be performed---both reliably and inexpensively---in the measurement space $\mathbb{R}^{m}$ (since $m\ll n$) \cite{hegde2008random}. In fact, it is even possible to reconstruct the original signal $x$ given only $\chi=\Phi_{l,u}x$ (and the knowledge of $\mathbb{M}$ and $\Phi_{l,u}$)~\cite{shah2011iterative,iwen2013approximation}.

In this context, two important cases are worth mentioning in regards to Theorem~\ref{thm:(Near-Isometry) rch preserved}. First, consider the case where the manifold $\M$ is concentrated in or around an $m$-dimensional subspace of $\R^{n}$, and suppose $\Phi_{l,u}$ is chosen as an orthogonal projection onto an orthogonal basis for that subspace. Such optimal projections are the topic of classical Principal Component Analysis (PCA). In this case, the $m$ nonzero singular values of $\Phi$ will exactly equal $1$, and if $l \approx 1$, Theorem~\ref{thm:(Near-Isometry) rch preserved} guarantees a lower bound on $\mbox{rch}\left(\Phi_{l,u}(\M)\right)$ that is approximately equal to $\mbox{rch}(\M)$.

As a second case, suppose $\Phi$ is generated randomly as an $m \times n$ matrix with $m < n$. Such matrices are commonly used as tool for dimensionality reduction in the field of Compressive Sensing~\cite{donoho2006compressed,candes2008restricted,cande2008introduction}. Consider, in particular, the situation where $\Phi$ is populated with independent and identically distributed Gaussian random variables, each with zero mean and variance $1/m$. In this situation, it is known that assuming
\begin{equation}
m\gtrsim\frac{k}{\Delta_{l,u}}\log\left[\frac{\left(\mbox{vol}_{k}(\M)\right)^{\frac{1}{k}}}{\mbox{rch}(\M)}\right],\label{eq:RIP}
\end{equation}
then with high probability (\ref{eq:iso}) will hold~\cite{eftekhari2015new} (see also~\cite{clarkson2008tighter}).\footnote{We have somewhat simplified the right hand side above for clarity, hence the sign $\gtrsim$. Exact details may be found in \cite{eftekhari2015new}.} In this case, one may indeed achieve a near-isometric embedding of the manifold with $l \approx u \approx 1$ and guarantee that the reach of the manifold does not collapse. However, the singular values of $\Phi$ will cluster around $\sqrt{n/m}$. Consequently, the lower bound on $\mbox{rch}\left(\Phi_{l,u}(\M)\right)$ will be lower than in the first case: around $\sqrt{m/n} \cdot \mbox{rch}(\M)$.

\bibliographystyle{plain}
\bibliography{References}

\appendix

\section{Proof of Proposition \ref{lem:ellipsoidreach}\label{sec:Proof-of-Lemma ellipsoid} \\ (Reach of an ellipsoid)}

% For the upper bound on the reach, we consider two points at each end of the minor axis of the ellipsoid. Each point has a normal vector that intersects the center of the ellipsoid, a distance of $r_n$ away. This gives an upper bound on reach of $r_n$.

To establish a lower bound on the reach, we rely on Blaschke's rolling theorem, as stated in~\cite{chazal2004stability}.

\begin{theorem}
\textbf{\label{thm:rolling}\emph{(Blaschke's rolling theorem~\cite{chazal2004stability})}}
Let $V \subset \mathbb{R}^n$ be a relatively compact convex open set. Assume that the boundary of $V$, $\partial V$, is a $C^2$ manifold. Assume that the maximum $K$ of the principal curvatures at any point of $\partial V$ is finite. Then, for all $0 < \epsilon < 1/K$ the Euclidean ball of radius $\epsilon$ can roll freely on $\partial V$ in the interior of $V$. More precisely, for all $x \in \partial V$, the ball of radius $\epsilon$ which is tangent to $\partial V$ at $x$ 
%and whose center point is inside $\partial V$ is contained in $V$ and
 has only $x$ as an intersection point with $\partial V$.
\end{theorem}

Letting $V$ denote the interior of our ellipsoid (so that $\partial V = \M$), the problem of bounding $\mbox{rch}(\M)$ reduces to that of bounding the maximum $K$ of the principal curvatures at any point of $\M$. Theorem~\ref{thm:rolling} then implies that the open normal bundle of $\mathbb{M}$ of radius $r$ is embedded in $\mathbb{R}^{n}$ for all $r<1/K$. Therefore, $\mbox{rch}(\M) \ge 1/K$.

To bound $K$, we consider the following argument. Without loss of generality, we suppose the ellipsoid is centered at the origin and aligned with the canonical axes, so that points $x \in \M$ are defined by the equation
\[
\sum_{i=1}^{n}\frac{x_{i}^{2}}{r_{i}^{2}}=1,
\]
for $r_{1}\ge\cdots\ge r_{n} > 0$. Equivalently, $x^{T}Dx=1$ with $D\in\mathbb{R}^{n\times n}$ the diagonal matrix containing $\{r_{i}^{-2}\}$.

Next, we consider a rigid transformation (which leaves reach unchanged) that shifts and rotates the ellipsoid as follows: for an arbitrary point $b \in \M$, we shift the point $b$ to the origin and rotate the ellipsoid so that, at the origin, it is orthogonal to $e_{n}$, the $n$th canonical vector in $\mathbb{R}^n$. More precisely, we consider the transformation $x=Ry+b$ for some unitary rotation matrix $R\in\mathbb{R}^{n\times n}$ to be defined below. Under this change of variables,
\[
1 =(Ry+b)^{T}D(Ry+b) =y^{T}R^{T}DRy+2b^{T}DRy+b^{T}Db.
\]
Since $b \in \M$, we have
\begin{equation}
b^{T}Db=1,\label{eq:f1}
\end{equation}
and the transformed ellipsoid becomes
\[
y^{T}R^{T}DRy+2b^{T}DRy=0,
\]
which passes through the origin. Ignoring the quadratic terms, we find that $2b^{T}DRy=0$. Therefore, the normal direction at the origin is $DR^{T}b$. We choose $R$ such that the normal vector is aligned with the last coordinate $e_n$, i.e.,
\begin{equation}
DR^{T}b=-\|DR^{T}b\|_{2}\cdot e_{n}.\label{eq:f2}
\end{equation}
With this choice, the ellipsoid is defined by the equation
\begin{equation}
y_{n}=\frac{y^{T}R^{T}DRy}{2\|DR^{T}b\|_{2}},\label{eq:ellips}
\end{equation}
which now passes through the origin and has the normal vector of $e_{n}$. Let us decompose $R^{T}DR$ according to the index $n$, i.e., we
let
\[
D^{'}=\left[\begin{array}{cc}
D{}_{11}^{'} & D'_{12}\\
D_{12}^{'T} & D'_{22}
\end{array}\right]:=R^{T}DR,
\]
for short, with
\[
D'_{11}\in\mathbb{R}^{(n-1)\times(n-1)},\quad D'_{12}\in\mathbb{R}^{n-1},\quad D'_{22}\in\mathbb{R}.
\]
Then, we can write (\ref{eq:ellips}) as
\[
2\|DR^{T}b\|_{2}\cdot y_{n}=y_{\backslash n}^{T}D'_{11}y_{\backslash n}+\left(2y_{\backslash n}^{T}D'_{12}\right)y_{n}+D'_{22}y_{n}^{2},
\]
where $y_{\backslash n}\in\mathbb{R}^{n-1}$ contains all the entries of $y$ except $y_{n}$. We now rotate the tangent plane $y_{n}=0$ by making the change of variables $z=Ly_{\backslash n}\in\mathbb{R}^{n-1}$ where the orthonormal matrix $L\in\mathbb{R}^{(n-1)\times(n-1)}$ diagonalizes $D'_{11}$. We arrive at
\[
2\|DR^{T}b\|_{2}\cdot y_{n}=z^{T}Ez+\left(2z^{T}F\right)y_{n}+D'_{22}y_{n}^{2},
\]
\[
E:=L^{T}D'_{11}L\in\mathbb{R}^{(n-1)\times(n-1)},\quad F:=L^{T}D'_{12}\in\mathbb{R}^{n-1},
\]
where $E$ is diagonal. Note that the first two terms (constant and
linear) of the Taylor expansion of $y_{n}$ in terms of $z$ are both
zero. The principal curvatures are the eigenvalues of the Hessian
$[\frac{\partial^{2}y_{n}}{\partial z_{j}\partial z_{j'}}]$ evaluated
at $z=0$. By taking derivatives of both sides, we may verify that
\[
\frac{\partial^{2}y_{n}}{\partial z_{j}\partial z_{j'}}|_{z=0}=\frac{E}{\|DR^{T}b\|_{2}}.
\]
Note that
\[
\left\Vert  \left. \frac{\partial^{2}y_{n}}{\partial z_{j}\partial z_{j'}}  \right\vert_{z=0} \right\Vert  \le\frac{\|E\|_{2}}{\|DR^{T}b\|_{2}}
 =\frac{\|L^{T}D'_{11}L\|}{\|DR^{T}b\|_{2}}
 \le\frac{\|D\|}{\|DR^{T}b\|_{2}}
 =\frac{r_{n}^{-2}}{\|DR^{T}b\|_{2}},
\]
where we recall that $r_{n}=\min_{i} r_{i}$. By design, $b^{T}Db=\sum b_{i}^{2}r_{i}^{-2}=1$ and consequently we note that
\begin{align*}
\min_{b}\frac{\|DR^{T}b\|_{2}}{\sqrt{b^{T}Db}} & =\min_{b}\frac{\|DR^{T}b\|_{2}}{\|D^{\frac{1}{2}}b\|_{2}}\\
 & =\min_{c}\frac{\|DR^{T}D^{-\frac{1}{2}}c\|_{2}}{\|c\|_{2}}\\
 & =\sqrt{\lambda_{min}\left[\left(D^{-\frac{1}{2}}RD^{\frac{1}{2}}\right)D\left(D^{\frac{1}{2}}R^{T}D^{-\frac{1}{2}}\right)\right]}\\
 & \ge r_{1}^{-1}\sqrt{\lambda_{min}\left[\left(D^{-\frac{1}{2}}RD^{\frac{1}{2}}\right)\left(D^{\frac{1}{2}}R^{T}D^{-\frac{1}{2}}\right)\right]}\\
 & =r_{1}^{-1}\sqrt{\lambda_{min}\left[D^{-\frac{1}{2}}RDR^{T}D^{-\frac{1}{2}}\right]}\\
 & =r_{1}^{-1}\sqrt{\lambda_{min}\left[D^{-1}RDR^{T}\right]}\\
 & =r_{1}^{-1}\sqrt{\lambda_{min}\left[D^{-1}RD\right]}\\
 & =r_{1}^{-1}\sqrt{\lambda_{min}\left[R\right]}\\
 & =r_{1}^{-1},
\end{align*}
where the sixth and eighth lines hold because the spectrum of a matrix is unchanged under a similarity transform. In the fourth line, we used the fact that $D\succcurlyeq r_{1}^{-2}\cdot I_{n}$ because $r_{1}=\max_{i} r_{i}$. Overall, we arrive at
\[
\left\Vert \frac{\partial^{2}y_{n}}{\partial z_{j}\partial z_{j'}}|_{z=0}\right\Vert \le\frac{r_{n}^{-2}}{r_{1}^{-1}}=\frac{r_{1}}{r_{n}^{2}}.
\]
That is, the largest principal curvature of an ellipsoid is bounded above by $K = r_{1}/r_{n}^{2}$. In fact, this is achieved at each endpoint of the major axis of the ellipsoid, which means that the largest principal curvature equals $r_{1}/r_{n}^{2}$ exactly. From the rolling theorem, we conclude that the reach of the ellipsoid is lower bounded by $r_{n}^{2}/r_{1}$.

In fact, Blaschke's rolling theorem is tight (see~\cite{howard1999blaschke}) in that---in the parlance of Theorem~\ref{thm:rolling}---the Euclidean ball of radius $1/K$ cannot roll freely on $\partial V$ in the interior of $V$. Since we have established above that $K = r_{1}/r_{n}^{2}$ exactly, it follows that the reach of the ellipsoid is also upper bounded bounded by $r_{n}^{2}/r_{1}$, and therefore, must equal exactly $r_{n}^{2}/r_{1}$.

\section{Proof of Proposition \ref{lem:easy}\label{sec:Proof-of-Lemma easy} \\ (Basic Properties)}

We prove only the claims concerning dimension and volume, as the others are self-evident. We also set $\Phi(\cdot) = \Phi_{l,u}(\cdot)$ for short.

Let $x \in \M$ be arbitrary. Fix a tangent vector $v\in\T_{x}\M$, and consider a curve $\gamma:[-1,1]\rightarrow\M$ such that
\begin{equation}
\gamma(0)=x,\qquad\frac{d\gamma}{dt}(0)=v.\label{eq:proof setup}
\end{equation}
Note that $\Phi\circ\gamma:[-1,1]\rightarrow\Phi(\M)$ is a curve on $\Phi(\M)$ that passes through  $\Phi(x)=\Phi(\gamma(0))$ with the tangent direction
\begin{equation}
\frac{d\left(\Phi\circ\gamma\right)}{dt}(0)=D\Phi(\gamma(0))\cdot\frac{d\gamma}{dt}(0)=D\Phi(x)\cdot v,\label{eq:cmp 1}
\end{equation}
where $D\Phi(x)\in\mathbb{R}^{m\times n}$ is the Jacobian of $\Phi(\cdot)$ at $x\in\mathbb{R}^{n}$. On the other hand,
\begin{align}
\left\Vert \frac{d\left(\Phi\circ\gamma\right)}{dt}(0)\right\Vert _{2} & =\lim_{t\rightarrow0}\frac{\left\Vert \Phi(\gamma(t))-\Phi(\gamma(0))\right\Vert _{2}}{t}\nonumber \\
 & \ge l\cdot\lim_{t\rightarrow0}\frac{\left\Vert \gamma(t)-\gamma(0)\right\Vert _{2}}{t}\qquad\mbox{(see (\ref{eq:near iso}))}\nonumber \\
 & =l\cdot\left\Vert \frac{d\gamma}{dt}(0)\right\Vert _{2}\nonumber \\
 & =l\cdot\left\Vert v\right\Vert _{2}>0,
\qquad \mbox{(see \eqref{eq:proof setup})} \label{eq:cmp 2}
\end{align}
where the second to last line holds because the derivative exists. Comparing (\ref{eq:cmp 1}) with (\ref{eq:cmp 2}), we conclude that the Jacobian $D\Phi(x):\T_{x}\M\rightarrow\T_{\Phi(x)}\Phi(\M)$ is an injective map. Since any tangent vector to $\Phi(\M)$ at $\Phi(x)$ can be written as in (\ref{eq:cmp 1}) (for some curve $\gamma(\cdot)$), $D\Phi(x)$ is in fact a bijective map.
% (when restricted to $\T_{x}\M$).
Since the choice of $x\in\M$ was arbitrary, we conclude that $\mbox{dim}(\Phi(\M))=\dim(\M)=k$.

In fact, a matching upper bound for (\ref{eq:cmp 2}) exists and we have that
\begin{equation}
l\|v\|_{2}\le\left\Vert D\Phi(x)\cdot v\right\Vert _{2}\le u\|v\|_{2},\qquad\forall v\in\T_{x}\M,\quad\forall x\in\M.\label{eq:l preserve tangent}
\end{equation}
That is, $\Phi(\cdot)$ nearly preserves the lengths of tangent vectors of $\M$. To prove the last claim in Proposition \ref{lem:easy}, note that
\[
\mbox{vol}_{k}\left(\M\right)=\int_{x\in\M}\mbox{vol}_{k}(dx),
\]
where $\mbox{vol}_{k}(dx)$ is the volume of the parallelotope formed by columns of
\[
dx\in\overset{k\:\mbox{times}}{\overbrace{\T_{x}\M\times\cdots\times\T_{x}\M}}\subset\mathbb{R}^{n\times k}.
\]
Similarly,
\[
\mbox{vol}_{k}\left(\Phi\left(\M\right)\right)=\int_{y\in\Phi(\M)}\mbox{vol}_{k}(dy),
\]
where
\[
dy\in\overset{k\:\mbox{times}}{\overbrace{\T_{\Phi(x)}\Phi\left(\M\right)\times\cdots\times\T_{\Phi(x)}\Phi\left(\M\right)}}\subset\mathbb{R}^{m\times k}.
\]
From  \eqref{eq:iso}, recall that $\Phi(\cdot)$ is injective on $\M$. Then, with the change of variables $y=\Phi(x)$, we find that
\begin{equation}
\mbox{vol}_{k}\left(\Phi\left(\M\right)\right)=\int_{x\in\M}\mbox{vol}_{k}\left(D\Phi(x)\cdot dx\right).\label{eq:volume relation}
\end{equation}
Note that
\[
\mbox{vol}_{k}\left(D\Phi(x)\cdot dx\right)=\sqrt{\det\left[dx^{T}\cdot D\Phi(x)^{T}D\Phi(x)\cdot dx\right]}.
\]
On the other hand, by (\ref{eq:l preserve tangent}), it holds that
\[
l^{2}\cdot dx^{T}\cdot dx\preccurlyeq dx^{T}\cdot D\Phi(x)^{T}D\Phi(x)\cdot dx\preccurlyeq u^{2}\cdot dx^{T}\cdot dx,
\]
and therefore
\[
l^{k}\cdot\sqrt{\det\left[dx^{T}\cdot dx\right]}\le\sqrt{\det\left[dx^{T}\cdot D\Phi(x)^{T}D\Phi(x)\cdot dx\right]}\le u^{k}\cdot\sqrt{\det\left[dx^{T}\cdot dx\right]},
\]
which reduces to
\[
l^{k}\cdot\mbox{vol}_{k}(dx)\le\mbox{vol}_{k}\left(D\Phi(x)\cdot dx\right)\le u^{k}\cdot\mbox{vol}_{k}(dx).
\]
Substituting the bound above into (\ref{eq:volume relation}) yields
\[
l^{k}\cdot\mbox{vol}_{k}(\M)\le\mbox{vol}_{k}\left(\Phi\left(\M\right)\right)\le u^{k}\cdot\mbox{vol}_{k}(\M),
\]
which completes the proof of Proposition \ref{lem:easy}.

\section{Proof of Proposition \ref{lem:key exact iso}\label{sec:Proof-of-Lemma key exact iso}\\(Tangent Bundle Under Exact Isometry)}

Fix $p,q\in\M$ and unit-length vectors  $e_p\in\T_{p}\M$ and $e_q\in\T_{q}\M$. Let $\theta_p,\theta_q:[-1,1]\rightarrow\M$ be
%\emph{unit speed}\footnote{$\theta(\cdot)$ is a unit-speed curve if $\|\frac{d\theta}{dt}(\cdot)\|_2=1$.}
geodesic curves  that pass through $p$  and $q$, respectively, such that
\[
\theta_p(0)=p,\qquad e_p=\theta'_p(0)=\lim_{t\rightarrow0}\frac{\theta_p(t)-\theta_p(0)}{t},
\]
\begin{equation}\label{eq:passing through}
\theta_q(0)=q,\qquad e_q=\theta_q'(0).
\end{equation}
For fixed $t>0$, we consider $\theta_p(t)-p$ on one hand and $\theta_q(t)-q$ on the other hand. We set $\Phi(\cdot) = \Phi_{l,u}(\cdot)$ for short. Because $\Phi(\cdot)$ is an exact isometry on $\M$, we have for any $t > 0$ that
\begin{align*}
& \left\langle \Phi\left(\theta_p(t)\right)-\Phi(p),\Phi\left(\theta_q(t)\right)-\Phi(q)\right\rangle \\
&= \left\langle \Phi\left(\theta_p(t)\right) , \Phi\left(\theta_q(t)\right) \right \rangle - \left\langle \Phi(p) , \Phi\left(\theta_q(t)\right) \right \rangle - \left\langle \Phi\left(\theta_p(t)\right) , \Phi(q) \right \rangle + \left\langle \Phi(p) , \Phi(q) \right \rangle \\
&= \frac{1}{2}\| \Phi\left(\theta_p(t)\right) - \Phi(q)\|^2 + \frac{1}{2}\| \Phi(p) - \Phi\left(\theta_q(t)\right) \|^2 \\ & \quad\quad - \frac{1}{2}\| \Phi\left(\theta_p(t)\right) - \Phi\left(\theta_q(t)\right)\|^2 - \frac{1}{2}\| \Phi(p) - \Phi(q)\|^2 \\
&= \frac{1}{2}\| \theta_p(t) - q\|^2 + \frac{1}{2}\| p - \theta_q(t) \|^2 - \frac{1}{2}\| \theta_p(t) - \theta_q(t)\|^2 - \frac{1}{2}\| p - q\|^2 \\
&= \left\langle \theta_p(t), \theta_q(t) \right \rangle - \left\langle p , \theta_q(t) \right \rangle - \left\langle \theta_p(t) , q \right \rangle + \left\langle p, q \right \rangle \\
&= \left\langle \theta_p(t)-p,\theta_q(t)-q\right\rangle,
\end{align*}
where the second and fourth equalities follow from the polarization identity. Hence,
\begin{align}
\left\langle
D\Phi(p) \cdot e_p,D\Phi(q)\cdot e_q \right\rangle
&=
\left\langle \lim_{t\rightarrow 0} \frac{\Phi\left( \theta_p(t)\right)- \Phi \left(p\right)}{t} , \lim_{t\rightarrow 0} \frac{\Phi\left( \theta_q(t)\right)- \Phi \left(q\right)}{t} \right\rangle
\nonumber\\
& =
\lim_{t\rightarrow 0} \frac{\left\langle \Phi\left( \theta_p(t)\right)- \Phi \left(p\right), \Phi\left( \theta_q(t)\right)- \Phi \left(q\right) \right\rangle}{t^2} \nonumber\\
& = \lim_{t\rightarrow 0} \frac{\left\langle  \theta_p(t)- p, \theta_q(t)- q \right\rangle}{t^2}\nonumber\\
&=
\left\langle \lim_{t\rightarrow 0} \frac{\theta_p(t)- p}{t} , \lim_{t\rightarrow 0} \frac{\theta_q(t)-q}{t} \right\rangle
\nonumber\\
& = \left\langle
e_p,e_q
\right\rangle.
\label{eq:inn prod iso}
\end{align}
Recalling the definition of principal angle, we then note that
\begin{align*}
\cos\left(\angle\left[\T_{\Phi(p)}\Phi(\M),
\T_{\Phi(q)}\Phi(\M)
\right]\right) &
 =\min_{
 e_{\Phi,p}\in\T_{\Phi(p)}\Phi(\M)
}
 \,\,
\max_{
 e_{\Phi,q}\in\T_{\Phi(q)}\Phi(\M)
}
\cos \left(\angle\left[
 e_{\Phi,p},
e_{\Phi,q}
 \right]\right)\\
 &
 = \min_{
 e_{p}\in\T_{p}\M
}
 \,\,
\max_{
 e_{q}\in\T_{q}\M
}
\cos \left(\angle\left[
 D\Phi(p) \cdot e_{p},
D\Phi(q) \cdot e_{q}
 \right]\right)\\
 &
= \min_{
 e_{p}\in\T_{p}\M
}
 \,\,
\max_{
 e_{q}\in\T_{q}\M
}
\frac{\left\langle D\Phi(p) \cdot e_p
,
D\Phi(q)\cdot e_q \right\rangle}{\left\|
D\Phi(p) \cdot e_p
\right\|_2
\cdot
\left\|
D\Phi(q) \cdot e_q
\right\|_2
}  \nonumber\\
& =
\min_{
 e_{p}\in\T_{p}\M
}
 \,\,
\max_{
 e_{q}\in\T_{q}\M
}
\frac{\left\langle  e_p
,
 e_q \right\rangle}{\left\|
e_p
\right\|_2
\cdot
\left\|
 e_q
\right\|_2
} \qquad \mbox{(see
\eqref{eq:l preserve tangent}  and
\eqref{eq:inn prod iso})}\\
& =
\cos \left(
\angle
\left[
\T_p\M, \T_q\M
\right]
\right).
\end{align*}
The second line above holds because $D\Phi(p):\T_{p}\M\rightarrow\T_{\Phi(p)}\Phi(\M)$ is a bijective map (see (\ref{eq:l preserve tangent})). This completes the proof of Proposition \ref{lem:key exact iso}.\footnote{For subspaces $\mathbb{A},\mathbb{B}$ and their orthogonal complements $\mathbb{A}^\perp,\mathbb{B}^\perp$, it holds that $\angle[\mathbb{A},\mathbb{B}]=\angle[\mathbb{A}^{\perp},\mathbb{B}^{\perp}]$. Therefore, a similar claim to that in Proposition \ref{lem:key exact iso} holds for normal spaces under an exact isometry. }

\section{Proof of Proposition \ref{lem:key near iso}\label{sec:Proof-of-Lemma key near iso}
\\ (Tangent Bundle Under Linear Near-Isometry)}

For short, we will use the notation $\Phi$ instead of $\Phi_{l,u}$, the matrix representation of the linear map $\Phi_{l,u}(\cdot)$. Fix $p,q\in\M$, as well as $e_p\in\T_{p}\M$ and $e_q\in\T_q\M$. Let $\theta_p:[-1,1]\rightarrow\M$ and  $\theta_q:[-1,1]\rightarrow\M$ be \emph{unit speed}\footnote{$\theta(\cdot)$ is a unit-speed curve if $\|\theta'(\cdot)\|_2=1$.}
geodesic curves passing through $p$ and $q$, respectively, such that \eqref{eq:passing through} holds. For some sufficiently small $\delta>0$, it holds that
\[
e_p=\theta_p'(0)\approx\frac{\theta_p(\delta)-\theta_p(0)}{\delta},
\qquad
e_q=\theta_q'(0)\approx\frac{\theta_q(\delta)-\theta_q(0)}{\delta}.
\]
More concretely,
\begin{align}
\left\Vert \theta_p'(0)-\frac{\theta_p(\delta)-\theta_p(0)}{\delta}\right\Vert_{2}
& =\frac{1}{\delta}\left\Vert \int_{0}^{\delta}\int_{0}^{\alpha} \theta_p''(\beta)\,d\beta d\alpha\right\Vert _{2}
\qquad\left(\theta''(\beta)=\frac{d^{2}\theta}{d\beta^{2}}(\beta)\right)
\nonumber\\
 & \le\frac{1}{\delta}\cdot\int_{0}^{\delta}\int_{0}^{\alpha}\left\Vert \theta_p''(\beta)\right\Vert _{2}\,d \beta d\alpha
 \nonumber\\
 & \le\frac{1}{\delta\cdot\mbox{rch}(\M)}\cdot\int_{0}^{\delta}\int_{0}^{\alpha}\,d\beta d\alpha\qquad\left(\mbox{Sec.~\ref{sec:Necessary-Concepts}: }\left\Vert \theta''(\beta)\right\Vert _{2}\le\frac{1}{\mbox{rch}(\M)}\right)
\nonumber \\
 & =\frac{\delta}{2\cdot\mbox{rch}(\M)},
\label{eq:der chord err p}
\end{align}
and similarly
\begin{equation}
\left\Vert \theta_q'(0)-\frac{\theta_q(\delta)-\theta_q(0)}{\delta}\right\Vert_{2}
\le
\frac{\delta}{2\cdot\mbox{rch}(\M)}.
\label{eq:der chord err q}
\end{equation}
To prove Proposition \ref{lem:key near iso}, we proceed as follows. First, we approximate $\langle e_p,e_q\rangle$ with a simpler quantity by noting that
\begin{align}
& \langle e_p,e_q \rangle
 = \left\langle
e_p - \frac{\theta_p(\delta)-\theta_p(0)}{\delta},
e_q
\right\rangle \nonumber\\
& +
\left\langle
\frac{\theta_p(\delta)-\theta_p(0)}{\delta},
e_q - \frac{\theta_q(\delta)-\theta_q(0)}{\delta}
\right\rangle +
\left\langle
\frac{\theta_p(\delta)-\theta_p(0)}{\delta},
\frac{\theta_q(\delta)-\theta_q(0)}{\delta}
\right\rangle.
\end{align}
The first two terms on the right hand side above are small thanks to \eqref{eq:der chord err p} and \eqref{eq:der chord err q}. Specifically,
\begin{align}
& \left|
\langle e_p,e_q \rangle-
\left\langle
\frac{\theta_p(\delta)-\theta_p(0)}{\delta},
\frac{\theta_q(\delta)-\theta_q(0)}{\delta}
\right\rangle
\right|\nonumber\\
& \le \left|
\left\langle
e_p - \frac{\theta_p(\delta)-\theta_p(0)}{\delta},
e_q
\right\rangle
\right|+
\left|
\left\langle
\frac{\theta_p(\delta)-\theta_p(0)}{\delta},
e_q - \frac{\theta_q(\delta)-\theta_q(0)}{\delta}
\right\rangle
\right|\nonumber\\
& \le
\left\| e_p-\frac{\theta_p(\delta)-\theta_p(0)}{\delta} \right\|_2 \cdot \|e_q\|_2 \nonumber \\ & \quad\quad +
\left(\left\|
e_p
\right\|_2
+ \left\|
e_p- \frac{\theta_p(\delta)-\theta_p(0)}{\delta}
\right\|_2
 \right) \cdot
 \left\|
 e_q-\frac{\theta_q(\delta)-\theta_q(0)}{\delta}
 \right\|_2\nonumber\\
 & \le
 \frac{\delta}{2\cdot \mbox{rch}(\M)} +  \left(
1+ \frac{\delta}{2\cdot \mbox{rch}(\M)} \right)
\cdot \frac{\delta}{2\cdot \mbox{rch}(\M)}
\qquad
\left(
\|e_p\|_2=\|e_q\|_2=1,
\mbox{ see \eqref{eq:der chord err p}
 and \eqref{eq:der chord err q}}
\right)
\nonumber\\
& \le \frac{5\delta}{4\cdot \mbox{rch}(\M)}.\qquad
\left( \mbox{if }
\delta \le \mbox{rch}(\M)
\right)
\label{eq:amenable 1}
\end{align}
Similarly, we approximate $\langle \Phi e_p,\Phi e_q\rangle$ with a more technically amenable quantity:
\begin{equation}
\left|
\langle \Phi e_p,\Phi e_q \rangle-
\left\langle
\Phi \cdot \frac{\theta_p(\delta)-\theta_p(0)}{\delta},
\Phi \cdot \frac{\theta_q(\delta)-\theta_q(0)}{\delta}
\right\rangle
\right|
\le
\frac{5\|\Phi\|^2\delta}{4\cdot \mbox{rch}(\M)}. ~~~
\left( \mbox{if }
\delta \le \mbox{rch}(\M)
\right)
\label{eq:amenable 2}
\end{equation}
Above, $\|\Phi\|$ is the spectral norm of the  matrix $\Phi$. We now use the triangle inequality, \eqref{eq:amenable 1}, and \eqref{eq:amenable 2} to simplify the difference $\langle\Phi e_p,\Phi e_q \rangle - \langle e_p,e_q\rangle$ as follows:
\begin{align}
& \left|
\langle \Phi e_p, \Phi e_q \rangle
- \langle e_p, e_q \rangle
\right|
\nonumber\\
& \le
\left|
\left\langle
\Phi \cdot \frac{\theta_p(\delta)-\theta_p(0)}{\delta},
\Phi \cdot \frac{\theta_q(\delta)-\theta_q(0)}{\delta}
\right\rangle
-
\left\langle
\frac{\theta_p(\delta)-\theta_p(0)}{\delta},
\frac{\theta_q(\delta)-\theta_q(0)}{\delta}
\right\rangle
\right| \nonumber \\ & \quad + \frac{5 \left(\|\Phi\|^2+1\right) \cdot \delta}{4\cdot \mbox{rch}(\M)}.
\label{eq:stage3}
\end{align}
To control the difference on the right hand side above, we write that
\begin{align*}
&
\left\langle \Phi \cdot \frac{\theta_p(\delta)-\theta_p(0)}{\delta}, \Phi \cdot \frac{\theta_q(\delta)-\theta_q(0)}{\delta} \right\rangle  \\
&= \frac{1}{2\delta^2} \left( \| \Phi\left(\theta_p(\delta) - \theta_q(0)\right)\|_2^2 + \| \Phi\left(\theta_p(0) - \theta_q(\delta)\right) \|_2^2
 - \| \Phi\left( \theta_p(\delta) - \theta_q(\delta)\right)\|_2^2 - \| \Phi\left(\theta_p(0) - \theta_q(0)\right)\|_2^2 \right) \\
&\le \frac{1}{2\delta^2} \left( u^2 \| \theta_p(\delta) - \theta_q(0)\|_2^2 + u^2 \| \theta_p(0) - \theta_q(\delta) \|_2^2  - l^2 \| \theta_p(\delta) - \theta_q(\delta)\|_2^2 - l^2 \| \theta_p(0) - \theta_q(0)\|_2^2 \right)
\qquad
\mbox{(see \eqref{eq:near iso})}
\\
&= \frac{1}{2\delta^2} \left(\| \theta_p(\delta) - \theta_q(0)\|_2^2 + \| \theta_p(0) - \theta_q(\delta) \|_2^2 - \| \theta_p(\delta) - \theta_q(\delta)\|_2^2 - \| \theta_p(0) - \theta_q(0)\|_2^2\right) \\ & \qquad + \frac{1}{2\delta^2}\left((u^2-1)\| \theta_p(\delta) - \theta_q(0)\|_2^2 + (u^2-1)\| \theta_p(0) - \theta_q(\delta) \|_2^2 \right.\\
& \qquad \qquad  \left.
+ (1-l^2)\| \theta_p(\delta) - \theta_q(\delta)\|_2^2  + (1-l^2)\| \theta_p(0) - \theta_q(0)\|_2^2\right) \\
&\le \frac{1}{2\delta^2} \left(\| \theta_p(\delta) - \theta_q(0)\|_2^2 + \| \theta_p(0) - \theta_q(\delta) \|_2^2 - \| \theta_p(\delta) - \theta_q(\delta)\|_2^2 - \| \theta_p(0) - \theta_q(0)\|_2^2\right) \\ & \qquad + \frac{4\Delta_{l,u}}{2\delta^2}  \left(\| \theta_p(0) - \theta_q(0)\|_2 + 2\delta\right)^2
\qquad
\left(\Delta_{l,u} := \max\left[1-l^2,u^2-1\right],
~\mbox{triangle ineq.}
\right)
 \\
&= \frac{1}{\delta^2} \left\langle \theta_p(\delta)-\theta_p(0) , \theta_q(\delta)-\theta_q(0)\right\rangle + \frac{2\Delta_{l,u}}{\delta^2}  \left(\| \theta_p(0) - \theta_q(0)\|_2 + 2\delta\right)^2
\qquad \mbox{(polarization id.)}
\\
&= \left\langle \frac{\theta_p(\delta)-\theta_p(0)}{\delta}, \frac{\theta_q(\delta)-\theta_q(0)}{\delta} \right\rangle +  \frac{2\Delta_{l,u}}{\delta^2} \left(\| \theta_p(0) - \theta_q(0)\|_2 + 2\delta\right)^2.
\end{align*}
Above, the first equality follows from the polarization identity. A matching lower bound is obtained similarly:
\begin{align*}
&
\left\langle \Phi \cdot \frac{\theta_p(\delta)-\theta_p(0)}{\delta}, \Phi \cdot \frac{\theta_q(\delta)-\theta_q(0)}{\delta} \right\rangle  \\
&\ge \left\langle \frac{\theta_p(\delta)-\theta_p(0)}{\delta}, \frac{\theta_q(\delta)-\theta_q(0)}{\delta} \right\rangle - \frac{2\Delta_{l,u}}{\delta^2} \left(\| \theta_p(0) - \theta_q(0)\|_2 + 2\delta\right)^2.
\end{align*}
Substituting the upper and lower bounds above back into \eqref{eq:stage3}, we find that
\begin{align}
& \left|
\left\langle\Phi e_p,\Phi e_q\right\rangle
- \left\langle e_p,e_q \right\rangle
\right|\nonumber\\
&
\le \frac{2\Delta_{l,u}}{\delta^2}  \left(\| \theta_p(0) - \theta_q(0)\|_2 + 2\delta\right)^2  +
\frac{5 \left(\|\Phi\|^2+1\right) \cdot \delta}{4\cdot \mbox{rch}(\M)} \nonumber\\
&
\le \frac{4\Delta_{l,u}}{\delta^2} \cdot \| p - q\|_2^2  +  16\Delta_{l,u}+
\frac{5 \left(\|\Phi\|^2+1\right) \cdot \delta}{4\cdot \mbox{rch}(\M)}
\qquad
\left( (a+b)^2 \le 2a^2+2b^2,\,\,\forall a,b\in\mathbb{R} \right) \\
&
\le \frac{4\Delta_{l,u}}{\delta^2} \cdot \| p - q\|_2^2  +  16\Delta_{l,u}+
\frac{5 \left(\max(\|\Phi\|^2,u^2)+1\right) \cdot \delta}{4\cdot \mbox{rch}(\M)}.
\end{align}
In the last line above we introduce the term $\max(\|\Phi\|^2,u^2)$. If the upper bound in \eqref{eq:near iso} is achieved for any pair of points, then it is guaranteed that $\|\Phi\| \ge u$; however, if \eqref{eq:near iso} is loose, then $u^2$ may dominate this maximum. Now, we may choose
\[
\delta = \frac{4\cdot \mbox{rch}(\mathbb{M})  \sqrt{\Delta_{l,u}}}{5\left( \max(\|\Phi\|^2,u^2)+1 \right)},
\]
which is guaranteed to be less than or equal to $\mbox{rch}(\mathbb{M})$, as required, and we find that
\begin{equation}
\label{eq:delta used}
\left|
\left\langle\Phi e_p,\Phi e_q\right\rangle
- \left\langle e_p,e_q \right\rangle
\right|
\le
\frac{25}{4} \cdot \frac{\|p-q\|_2^2}{ \mbox{rch}(\mathbb{M})^2}  \cdot \left(\max(\|\Phi\|^2,u^2)+1 \right)^2
 +  17 \max\left(\Delta_{l,u},\sqrt{\Delta_{l,u}}\right).
\end{equation}
Equipped with this estimate, we argue that \begin{align*}
&\cos\left( \angle\left[
\T_{\Phi p}\Phi \M, \T_{\Phi q}\Phi{\M})
\right]
\right) \\
& = \min_{e_{\Phi,p}\in \T_{\Phi p}\Phi \M}
\,\,
\max_{e_{\Phi,q}\in \T_{\Phi q}\Phi \M}
\cos\left(
\angle \left[
e_{\Phi,p},e_{\Phi,q}
\right]
\right)\nonumber\\
& = \min_{e_p\in\T_{p}\M}\,\,
\max_{e_q\in\T_q\M}
\cos
\left(
\angle \left[
\Phi  e_p,\Phi e_q
\right]
\right)\nonumber\\
& = \min_{e_p\in\T_{p}\M}\,\,
\max_{e_q\in\T_q\M}
\frac{
\left\langle
\Phi e_p, \Phi e_q
\right\rangle
}{
\left\|
\Phi e_p
\right\|_2
\cdot
\left\|
\Phi e_q
\right\|_2
}\nonumber\\
&
\le
\min_{e_p\in\T_{p}\M}\,\,
\max_{e_q\in\T_q\M}
\frac{
\left\langle
e_p,  e_q
\right\rangle
+
\left|
\left\langle
\Phi e_p, \Phi e_q
\right\rangle
-
\left\langle
e_p,e_q
\right\rangle
\right|
}{ l\cdot  \|e_p\|_2 \cdot l \cdot \|e_q\|_2}
\qquad \mbox{(triangle inequality and \eqref{eq:l preserve tangent})}
\nonumber\\
& =
\min_{e_p\in\T_{p}\M}\,\,
\max_{e_q\in\T_q\M}
\frac{\langle e_p,e_q\rangle
+ \left|
\left\langle
\Phi e_p, \Phi e_q
\right\rangle
-
\left\langle
e_p,e_q
\right\rangle
\right|
}{l^2}
\qquad \left(
\|e_p\|_2=\|e_q\|_2=1
\right)\nonumber\\
& \le \left( \frac{1}{l^2} \cdot
\min_{e_p\in\T_{p}\M}\,\,
\max_{e_q\in\T_q\M}
\left\langle
e_p,e_q
\right\rangle \right)
+
\frac{25}{4l^2} \cdot \frac{\|p-q\|_2^2}{\mbox{rch}(\mathbb{M})^2}  \cdot \left(\max(\|\Phi\|^2,u^2)+1 \right)^2
\\ & \qquad  +  \frac{17 \max\left(\Delta_{l,u},\sqrt{\Delta_{l,u}}\right)}{l^2}
 \qquad
 \mbox{(see \eqref{eq:delta used})}
 \nonumber\\
& = {l^{-2}}\cdot\cos\left(\angle\left[\T_p\M,\T_q\M\right]\right) + \frac{25}{4l^2} \cdot \frac{\|p-q\|_2^2}{  \mbox{rch}(\mathbb{M})^2}  \cdot \left(\max(\|\Phi\|^2,u^2)+1 \right)^2 + \frac{17\max\left(\Delta_{l,u},\sqrt{\Delta_{l,u}}\right)}{l^2} \nonumber\\
& = \cos\left(\angle\left[\T_p\M,\T_q\M\right]\right) + \frac{1-l^{2}}{l^{2}}\cdot\cos\left(\angle\left[\T_p\M,\T_q\M\right]\right) \\ & \quad\quad + \frac{25}{4l^2} \cdot \frac{\|p-q\|_2^2}{  \mbox{rch}(\mathbb{M})^2}  \cdot \left(\max(\|\Phi\|^2,u^2)+1 \right)^2  + \frac{17\max\left(\Delta_{l,u},\sqrt{\Delta_{l,u}}\right)}{l^2} \nonumber \\
& \le \cos\left(\angle\left[\T_p\M,\T_q\M\right]\right) + \frac{1-l^{2}}{l^{2}} + \frac{25}{4l^2} \cdot \frac{\|p-q\|_2^2}{  \mbox{rch}(\mathbb{M})^2}  \cdot \left(\max(\|\Phi\|^2,u^2)+1 \right)^2 + \frac{17\max\left(\Delta_{l,u},\sqrt{\Delta_{l,u}}\right)}{l^2} \nonumber \\
& \le \cos\left(\angle\left[\T_p\M,\T_q\M\right]\right) + \frac{25}{4l^2} \cdot \frac{\|p-q\|_2^2}{\Delta_{l,u} \cdot  \mbox{rch}(\mathbb{M})^2}  \cdot \left(\max(\|\Phi\|^2,u^2)+1 \right)^2 + \frac{18\max\left(\Delta_{l,u},\sqrt{\Delta_{l,u}}\right)}{l^2}.
\end{align*}
A lower bound is obtained similarly:
\begin{align*}
& \cos\left( \angle\left[\T_{\Phi p}\Phi \M, \T_{\Phi q}\Phi{\M}\right]\right) \\
& \ge {u^{-2}}\cdot\cos\left(\angle\left[\T_p\M,\T_q\M\right]\right)-\frac{25}{4u^2} \cdot \frac{\|p-q\|_2^2}{ \mbox{rch}(\mathbb{M})^2}  \cdot \left(\max(\|\Phi\|^2,u^2)+1 \right)^2 -  \frac{17\max\left(\Delta_{l,u},\sqrt{\Delta_{l,u}}\right)}{u^2} \\
& = \cos\left(\angle\left[\T_p\M,\T_q\M\right]\right) - \frac{u^{2} - 1}{u^{2}}\cdot\cos\left(\angle\left[\T_p\M,\T_q\M\right]\right) \\ & \quad\quad -\frac{25}{4u^2} \cdot \frac{\|p-q\|_2^2}{\mbox{rch}(\mathbb{M})^2}  \cdot \left(\max(\|\Phi\|^2,u^2)+1 \right)^2 -  \frac{17\max\left(\Delta_{l,u},\sqrt{\Delta_{l,u}}\right)}{u^2} \\
& \ge \cos\left(\angle\left[\T_p\M,\T_q\M\right]\right) -\frac{25}{4u^2} \cdot \frac{\|p-q\|_2^2}{ \mbox{rch}(\mathbb{M})^2}  \cdot \left(\max(\|\Phi\|^2,u^2)+1 \right)^2 -  \frac{18\max\left(\Delta_{l,u},\sqrt{\Delta_{l,u}}\right)}{u^2} \\
& \ge \cos\left(\angle\left[\T_p\M,\T_q\M\right]\right) -\frac{25}{4l^2} \cdot \frac{\|p-q\|_2^2}{ \mbox{rch}(\mathbb{M})^2}  \cdot \left(\max(\|\Phi\|^2,u^2)+1 \right)^2 -  \frac{18\max\left(\Delta_{l,u},\sqrt{\Delta_{l,u}}\right)}{l^2},
\end{align*}
where the last line uses the fact that $l \le u$. This completes the proof of Proposition \ref{lem:key near iso}.

\section{Proof of Proposition \ref{lem:exact iso}\label{sec:proof of exact iso} \\ (Reach Under Exact Isometry)}

By our assumption,
\[
\left\Vert \Phi_{1,1}(x)-\Phi_{1,1}(y)\right\Vert _{2}=\|x-y\|,\qquad\forall x,y\in\M,
\]
and we wish to determine $\mbox{rch}(\Phi_{1,1}(\M))$ in relation to $\mbox{rch}(\M)$. We set $\Phi(\cdot) = \Phi_{1,1}(\cdot)$ for short throughout the proof.

Fix an arbitrary point $x\in\M$. Take another arbitrary point $y\in\M$ and let $\gamma(\cdot)$ be a unit-speed geodesic curve that passes through $x$ and $y$ so that
\begin{equation}
\gamma(0)=x,\quad\gamma(d)=y,
\label{eq:proof setup 2}
\end{equation}
for some $d>0$. Next, we consider $\Phi(x),\Phi(y)\in\Phi(\M)$ and fix a normal vector $w_{\Phi}\in\N_{\Phi(x)}\Phi(\M)$ with $\|w_{\Phi}\|_{2}=1$. For
$r_{\Phi}>0$ to be assigned later, we set
\begin{equation}
z_{\Phi}=\Phi(x)+r_{\Phi}\cdot w_{\Phi}.\label{eq:def of z eps}
\end{equation}
Our objective is to show that, for sufficiently small $r_{\Phi}$, $z_{\Phi}$ is closer to $\Phi(x)$ than any other point $\Phi(y)\in\Phi(\M)$. To that end, we write that
\begin{align}
& \left\Vert \Phi(y)-z_{\Phi}\right\Vert _{2}^{2}-r_{\Phi}^{2} \nonumber \\ & =\left\Vert \Phi(y)-z_{\Phi}\right\Vert _{2}^{2}-\left\Vert \Phi(x)-z_{\Phi}\right\Vert _{2}^{2}
\qquad \mbox{(see \eqref{eq:def of z eps})}
\nonumber \\
 & =\left\Vert \Phi\left(\gamma(d)\right)-z_{\Phi}\right\Vert _{2}^{2}-\left\Vert \Phi\left(\gamma(0)\right)-z_{\Phi}\right\Vert _{2}^{2}
\qquad \mbox{(see \eqref{eq:proof setup 2})}
 \nonumber \\
 & =\int_{0}^{d}\frac{d}{dt}\left\Vert \Phi\left(\gamma(t)\right)-z_{\Phi}\right\Vert _{2}^{2}\,dt\nonumber \\
 & =2\int_{0}^{d}\left\langle \Phi\left(\gamma(t)\right)-z_{\Phi},D\Phi(\gamma(t))\cdot\gamma'(t)\right\rangle \,dt\qquad\left(\gamma'(t):=\frac{d\gamma}{dt}(t)\right)\nonumber \\
 & =2\int_{0}^{d}\left\langle \Phi\left(\gamma(t)\right)-\Phi\left(\gamma(0)\right)-r_{\Phi}\cdot w_{\Phi},D\Phi(\gamma(t))\cdot\gamma'(t)\right\rangle \,dt\qquad\mbox{(see (\ref{eq:def of z eps}))}\nonumber \\
 & =2\int_{0}^{d}\left\langle \Phi\left(\gamma(t)\right)-\Phi\left(\gamma(0)\right),D\Phi(\gamma(t))\cdot\gamma'(t)\right\rangle \,dt-2r_{\Phi}\int_{0}^{d}\left\langle w_{\Phi},D\Phi(\gamma(t))\cdot\gamma'(t)\right\rangle \,dt\nonumber \\
 & =\left(2\int_{0}^{d}\left\langle \Phi\left(\gamma(t)\right)-\Phi\left(\gamma(0)\right),D\Phi(\gamma(t))\cdot\gamma'(t)\right\rangle \,dt\right)-2r_{\Phi}\left\langle w_{\Phi},\Phi\left(\gamma(d)\right)-\Phi\left(\gamma(0)\right)\right\rangle \nonumber \\
 & =\left(2\int_{0}^{d}\left\langle \Phi\left(\gamma(t)\right)-\Phi\left(\gamma(0)\right),D\Phi(\gamma(t))\cdot\gamma'(t)\right\rangle \,dt\right)-2r_{\Phi}\left\langle w_{\Phi},\Phi(y)-\Phi(x)\right\rangle .\label{eq:1-1}
\end{align}
Above, we used the fact that
\[
\frac{d}{dt}\left\Vert \Phi\left(\gamma(t)\right)-z_{\Phi}\right\Vert _{2}^{2}=2 \left\langle \Phi\left(\gamma(t)\right)-z_{\Phi},D\Phi(\gamma(t))\cdot\gamma'(t)\right\rangle ,
\]
where $D\Phi(\cdot)\in\mathbb{R}^{m\times n}$ is the Jacobian of $\Phi(\cdot)$. In order to deal with the two terms in the last line of (\ref{eq:1-1}), we invoke a variation of Proposition \ref{lem:key exact iso}  (which we state without proof).
\begin{lemma}
Take $p,q\in\M$ and fix $e\in\T_{p}\M$ with unit length. Then
\[
\left\langle \Phi(q)-\Phi(p),D\Phi(p)\cdot e\right\rangle =\left\langle q-p,e\right\rangle .
\]
Moreover,
\[
\angle\left[\Phi(q)-\Phi(p),\T_{\Phi(p)}\M\right]=\angle\left[q-p,\T_{p}\M\right],
\]
\[
\angle\left[\Phi(q)-\Phi(p),\N_{\Phi(p)}\M\right]=\angle\left[q-p,\N_{p}\M\right],
\]
where $\angle[q-p,\T_{p}\M]$ is the smallest angle between $q-p$ and all directions in the subspace $\T_{p}\M$. The other angles are defined similarly.
\end{lemma}

Thanks to the above result, there exists a unit length normal vector $w\in\N_x\M$ for which we can write the following concerning the two inner products in the last line of \eqref{eq:1-1}:\footnote{For the second identity, the existence of normal vector $w$ is guaranteed for the following reason. Set $\alpha=\angle[y-x,\N_x\M]=\angle[\Phi(y)-\Phi(x),\N_{\Phi(x)}\Phi(\M)]$. Then, by definition of the angle between subspaces, for any angle $\beta\ge \alpha$, there must exist some unit length vector in $\N_x\M$ (say $w$) such that $\beta=\angle[y-x,w]$.}
\[
\left\langle \Phi\left(\gamma(t)\right)-\Phi\left(\gamma(0)\right),D\Phi(\gamma(t))\cdot\gamma'(t)\right\rangle =\left\langle \gamma(t)-\gamma(0),\gamma'(t)\right\rangle ,
\]
\[
\left\langle w_{\Phi},\Phi(y)-\Phi(x)\right\rangle =\left\langle w,y-x\right\rangle.
\]
We are now in position to rewrite the last line of (\ref{eq:1-1}) and find that
\begin{align*}
\left\Vert \Phi(y)-z_{\Phi}\right\Vert _{2}^{2}-r_{\Phi}^{2} & =\left(2\int_{0}^{d}\left\langle \gamma(t)-\gamma(0),\gamma'(t)\right\rangle \,dt\right)-2r_{\Phi}\left\langle w,y-x\right\rangle .
\end{align*}
We set $z=x+r_{\Phi}\cdot w$ and trace back our steps in (\ref{eq:1-1}) to arrive at
\begin{align*}
\left\Vert \Phi(y)-z_{\Phi}\right\Vert _{2}^{2}-r_{\Phi}^{2} & =\left(2\int_{0}^{d}\left\langle \gamma(t)-\gamma(0),\gamma'(t)\right\rangle \,dt\right)-2r_{\Phi}\left\langle w,y-x\right\rangle \\
 & =\left\Vert y-z\right\Vert _{2}^{2}-r_{\Phi}^{2},
\end{align*}
or
\begin{equation}
\left\Vert \Phi(y)-z_{\Phi}\right\Vert _{2}^{2}=\left\Vert y-z\right\Vert _{2}^{2}.\label{eq:key in reach}
\end{equation}
By definition, as long as $r_{\Phi}<\mbox{rch}(\M)$, $z$ is closer to $x$ than it is to $y$, i.e.,
\[
r_{\Phi}<\mbox{rch}(\M)\Longrightarrow\left\Vert z-y\right\Vert _{2}^{2}>\left\Vert z-x\right\Vert _{2}^{2}=r_{\Phi}^{2}.
\]
Now, thanks to (\ref{eq:key in reach}), as long as $r_{\Phi}<\mbox{rch}(\M)$, $z_{\Phi}$ is closer to $\Phi(x)$ than it is to $\Phi(y)$, i.e.,
\[
r_{\Phi}<\mbox{rch}(\M)\Longrightarrow\left\Vert z_{\Phi}-\Phi(y)\right\Vert _{2}^{2}>\left\Vert z_{\Phi}-\Phi(x)\right\Vert _{2}^{2}=r_{\Phi}^{2}.
\]
Our choice of $x,y\in\M$ and $w_{\Phi}\in\N_{\Phi(x)}\Phi(\M)$ were arbitrary and therefore the reach of $\Phi(\M)$ is at least $\mbox{rch}(\M)$, i.e., $\mbox{rch}(\Phi(\M))\ge\mbox{rch}(\M)$. In the opposite direction, we find that $\mbox{rch}(\M)\ge \mbox{rch}(\Phi(\M))$, as the (inverse) map that takes $\Phi(\M)$ back to $\M$ too is obviously an isometry. This completes the proof of Proposition \ref{lem:exact iso}.

\section{Proof of Proposition \ref{fact:In-general,-reach} \label{sec:Proof-of-Proposition in general} \\ (Negative Result)}

\begin{figure}[t]
\centering
(a) \includegraphics[scale=0.25]{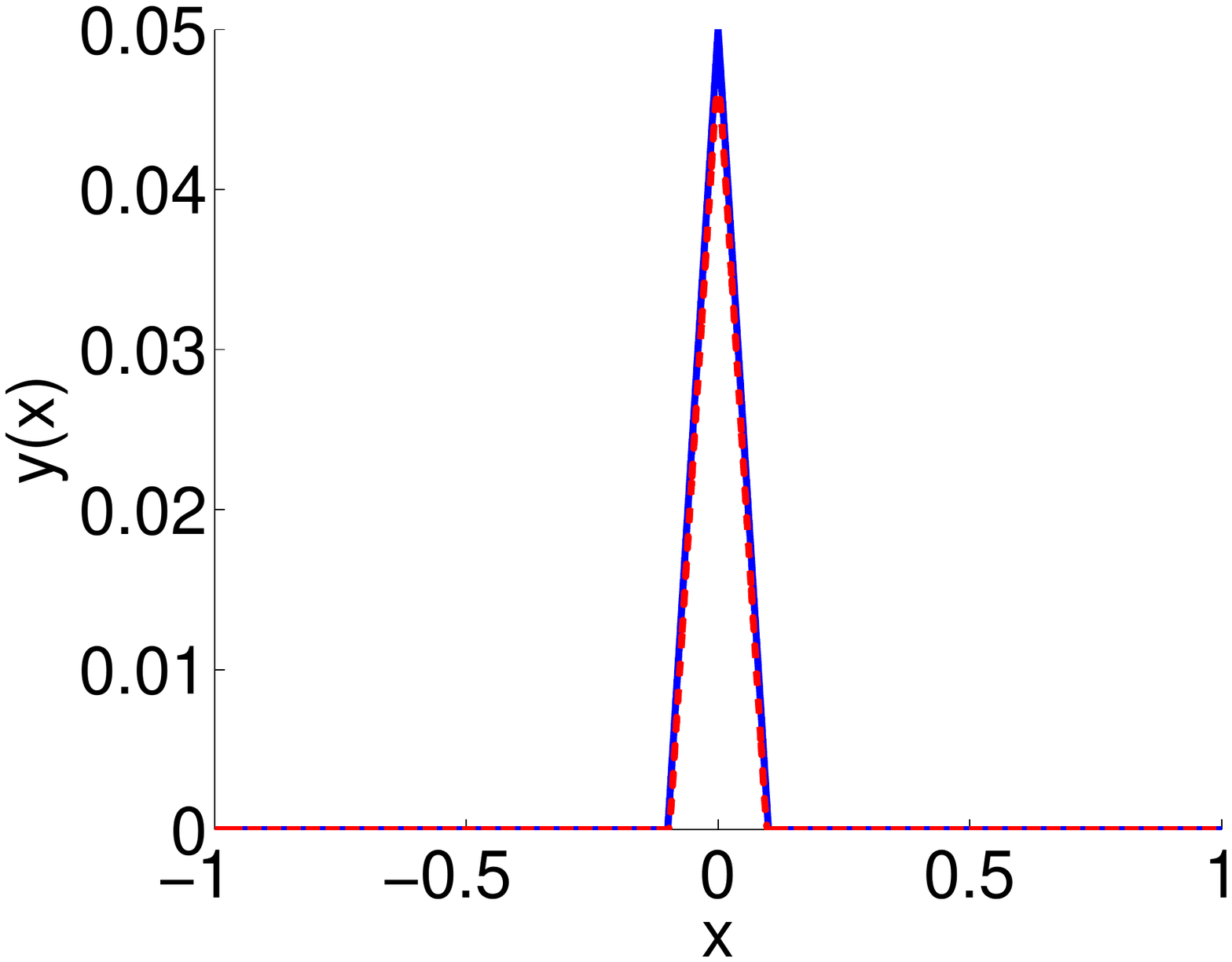} ~~~
(b) \includegraphics[scale=0.25]{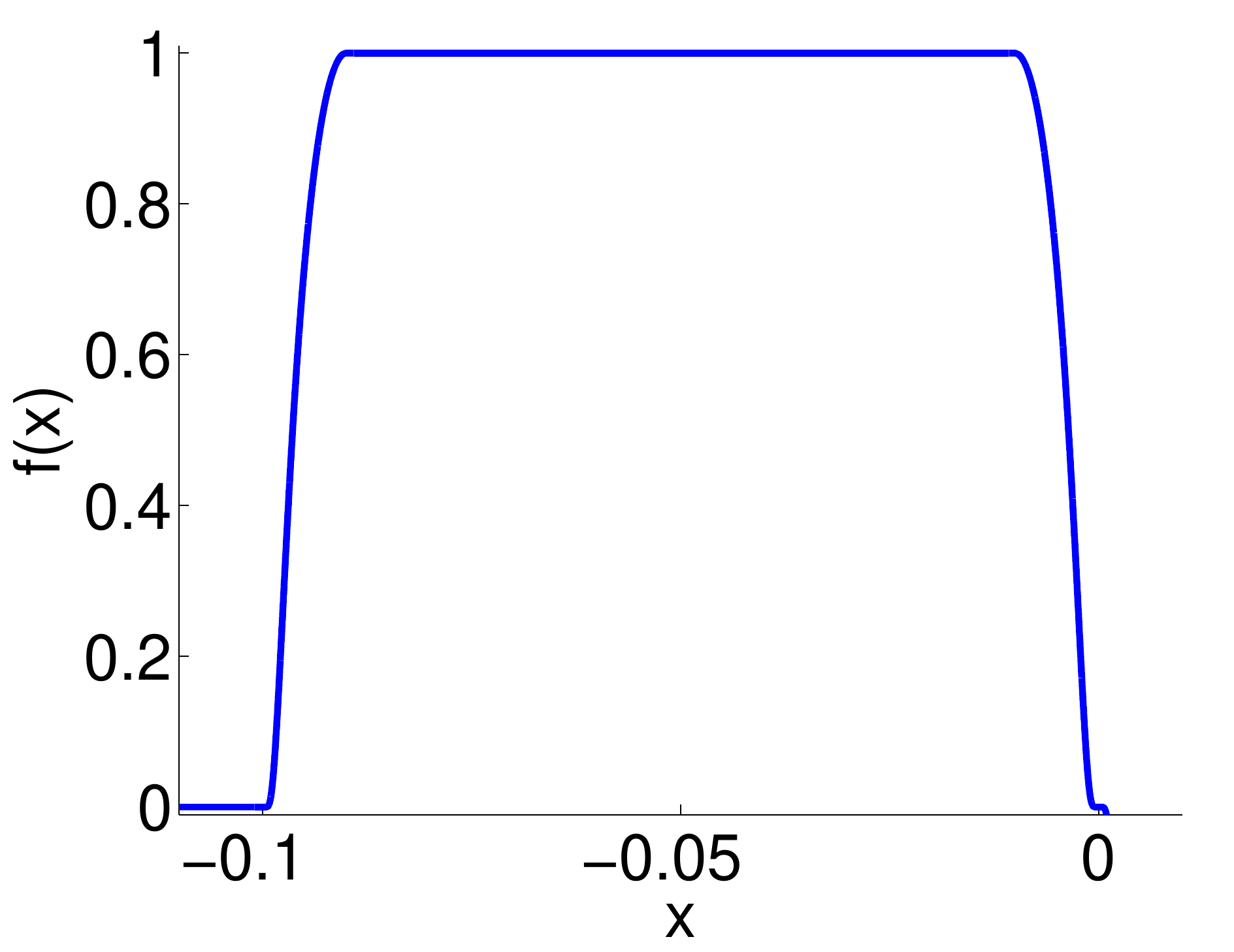} ~~~
\caption{\small\sl Setup in Appendix \ref{sec:Proof-of-Proposition in general}. (a) The solid blue curve shows a plot of $y_{\M_{\circ}}(x)$ as a function of $x$, while the dashed red curve shows a plot of $y_{\M_{*}}(x)$. (b) Plot of $f(x)$ for a narrow range of $x$. In all plots, $\delta = 0.1$, $c = 0.05$, and $\rho = 0.01$.}
\label{fig:1dcurves}
\end{figure}

Here, we establish Proposition \ref{fact:In-general,-reach} by constructing a suitable example. Suppose that $\M\subset\R^{2}$ is a part of the horizontal axis in $\R^{2}$. More specifically, we set
\begin{equation}
\M=\left\{ \left[\begin{array}{c}
x\\
y_{\M}(x)
\end{array}\right]\,:\,y_{\M}(x)=0,\quad|x|<1\right\} .\label{eq:M example}
\end{equation}
Note that $\mbox{rch}(\M)=\infty$. We next design a smooth manifold $\M_{*}\subset\mathbb{R}^{2}$ such that
\begin{itemize}
\item $\M_{*}$ is nearly-isometric to $\M$, and yet
\item $\mbox{rch}(\M_{*})\approx0\ll\infty=\mbox{rch}(\M)$.
\end{itemize}
To that end, we proceed in two steps. First, we place a triangle wave function at the origin. More specifically, given $0<c<\delta<1$, let
\begin{equation}
y_{\mathbb{M}_{\circ}}(x)=c\cdot\begin{cases}
1-\frac{|x|}{\delta} & |x|\le\delta,\\
0 & \delta\le|x|<1,
\end{cases}\label{eq:Mo}
\end{equation}
be the triangle wave. This specifies a manifold in $\mathbb{R}^{2}$ which we denote by $\M_{\circ}$:
\[
\M_{\circ}=\left\{ \left[\begin{array}{c}
x\\
y_{\M_{\circ}}(x)
\end{array}\right]\,:\,|x|<1\right\} \subset\mathbb{R}^{2}.
\]
The solid blue curve in Figure~\ref{fig:1dcurves}(a) shows a plot of $y_{\M_{\circ}}(x)$.
Note that $\M_{\circ}$ is not smooth because of the corners (cusps) at $[\pm\delta,0]^{T}$ and at $[0,c]^T$. This manifold is therefore not suited for our example because the map between $\M$ and $\M_{*}$ must be, by our assumption, smooth. To work around this issue, we simply smooth out the corners of $\M_{\circ}$ as follows. Given $0<\rho\ll\delta$ and when $x\in(0,1)$, let
\begin{equation}
f(x)=\begin{cases}
-\exp\left(1-\frac{1}{1-\left(\frac{x-\rho}{\rho}\right)^{2}}\right) & 0<x\le\rho,\\
-1 & \rho\le x\le\delta-\rho,\\
-\exp\left(1-\frac{1}{1-\left(\frac{x-(\delta-\rho)}{\rho}\right)^{2}}\right) & \delta-\rho\le x<\delta,\\
0 & \delta\le x,
\end{cases}\label{eq:def of f}
\end{equation}
and, when $x\in(-1,0)$, set
\[
f(x)=-f(-x),\qquad x<0.
\]
We also set $f(0)=0$. Closely related to the so-called \emph{bump function}, $f(\cdot)$ is an odd function compactly supported on $[-\delta,\delta]$ which is also smooth (and in fact infinitely differentiable). We now specify $\M_{*}$ as follows:
\begin{equation}\label{eq:def of M*}
\M_{*}=\left\{ \left[\begin{array}{c}
x\\
y_{\M_{*}}(x)
\end{array}\right]\,:\,y_{\M_{*}}(x)=\frac{c}{\delta}\cdot\int_{-\infty}^{x}f(z)\,dz,\quad|x|<1\right\} \subset\R^{2}.
\end{equation}
The dashed red curve in Figure~\ref{fig:1dcurves}(a) shows a plot of $y_{\M_{*}}(x)$. Note that $\M_{*}$ closely resembles $\M_{\circ}$ but, in contrast, is smooth. Furthermore, it is not difficult to verify that
\begin{equation}
\mbox{rch}\left(\M_{\circ}\right),\,\mbox{rch}\left(\M_{*}\right)\le\frac{\delta}{\sqrt{2}}.\label{eq:rch of M*}
\end{equation}
Indeed, for $\M_{\circ}$, the origin has a distance of at most $\delta/\sqrt{2}$ from both of the line segments starting at $[0,c]^{T}$ and ending at $[\pm\delta,0]^{T}$. Therefore, $\mbox{rch}(\M_{\circ})\le\delta/\sqrt{2}$. The same statement holds true for $\M_{*}$ since the graph of $\M_{*}$ is also symmetric about the vertical axis and lies beneath that of $\M_{\circ}$ (i.e., $0\le y_{\M_{*}}(\cdot)\le y_{\M_{\circ}}(\cdot)$), and lastly because the origin does not belong to $\M_*$.

{\bf Calculating the Isometry Constant of $\M_{*}$.} We next show that $\M_{*}$ is nearly-isometric to $\M$. First, from (\ref{eq:Mo}) and (\ref{eq:def of M*}), note that
\begin{equation}\label{eq:compare ders}
\frac{c}{\delta}\left|f(x)\right|
= \left|\frac{dy_{\M_{*}}}{dx}(x)\right|
\le
\left|\frac{dy_{\M_{\circ}}}{dx}(x)\right|,\qquad\forall x\notin\{0,\pm\delta\}.
\end{equation}
Then, from \eqref{eq:compare ders} and the fact that both $f(\cdot)$ and the derivative of $y_{\M_{\circ}}(\cdot)$ only change sign at the origin, we write the following whenever both $x_{1},x_{2}\in(-1,1)$ have the same sign:
\begin{align*}
\left|y_{\M_{*}}(x_{1})-y_{\M_{*}}(x_{2})\right| & =\frac{c}{\delta}\left|\int_{x_{2}}^{x_{1}}f(x)\,dx\right|\\
 & =\frac{c}{\delta}\int_{x_{2}}^{x_{1}}\left|f(x)\right|\,dx\\
 & \le\int_{x_{2}}^{x_{1}}\left|\frac{dy_{\M_{\circ}}}{dx}(x)\right|\,dx
\qquad \mbox{(see \eqref{eq:compare ders})}
 \\
 & =\left|\int_{x_{2}}^{x_{1}}\frac{dy_{\M_{\circ}}}{dx}(x)\,dx\right|\\
 & =\left|y_{\M_{\circ}}(x_{1})-y_{\M_{\circ}}(x_{2})\right|.
\end{align*}
That is, on each side of the origin, $\M_*$ is a \emph{contraction} of $\M_{\circ}$. This conclusion is in fact true everywhere. Indeed, for arbitrary $x_1,x_2\in(-1,1)$, we use the symmetry of both $\M_{\circ}$ and $\M_*$ to argue that
\begin{align*}
\left|y_{\M_{*}}(x_{1})-y_{\M_{*}}(x_{2})\right| & = \left|y_{\M_{*}}\left(|x_{1}|\right)-y_{\M_{*}}\left(|x_{2}|\right)\right|
\qquad \mbox{(by symmetry)}
\\
& \le\left|y_{\M_{\circ}}(x_{1})-y_{\M_{\circ}}(x_{2})\right|,\qquad\forall x_{1},x_{2}\in(-1,1),
\end{align*}
and, consequently,
\begin{equation}
1\le\frac{\left\Vert \left[\begin{array}{c}
x_{1}\\
y_{\M_{*}}(x_{1})
\end{array}\right]-\left[\begin{array}{c}
x_{2}\\
y_{\M_{*}}(x_{2})
\end{array}\right]\right\Vert _{2}}{\left\Vert \left[\begin{array}{c}
x_{1}\\
y_{\M}(x_{1})
\end{array}\right]-\left[\begin{array}{c}
x_{2}\\
y_{\M}(x_{2})
\end{array}\right]\right\Vert _{2}}\le\frac{\left\Vert \left[\begin{array}{c}
x_{1}\\
y_{\M_{\circ}}(x_{1})
\end{array}\right]-\left[\begin{array}{c}
x_{2}\\
y_{\M_{\circ}}(x_{2})
\end{array}\right]\right\Vert _{2}}{\left\Vert \left[\begin{array}{c}
x_{1}\\
y_{\M}(x_{1})
\end{array}\right]-\left[\begin{array}{c}
x_{2}\\
y_{\M}(x_{2})
\end{array}\right]\right\Vert _{2}},\quad\forall x_{1},x_{2}\in(-1,1).\label{eq:M* n Mo}
\end{equation}
(Above, we invoked the fact that $y_{\M}(x)=0$ for all $x$, so that the  denominators above all equal $|x_1-x_2|$.) In words, the isometry constant of $\M_{*}$ (as a mapping from $\M$) is always upper-bounded by the isometry constant of $\M_{\circ}$. We therefore focus on computing the latter quantity, as it is more convenient.

To that end, we use (\ref{eq:M example}) and (\ref{eq:Mo}) as follows. There are a few possibilities which we next study one by one. For a pair of points $x_{1},x_{2}$ far from the origin, we have that
\[
\frac{\left\Vert \left[\begin{array}{c}
x_{1}\\
y_{\M_{\circ}}(x_{1})
\end{array}\right]-\left[\begin{array}{c}
x_{2}\\
y_{\M_{\circ}}(x_{2})
\end{array}\right]\right\Vert _{2}}{\left\Vert \left[\begin{array}{c}
x_{1}\\
y_{\M}(x_{1})
\end{array}\right]-\left[\begin{array}{c}
x_{2}\\
y_{\M}(x_{2})
\end{array}\right]\right\Vert _{2}}=\frac{\left\Vert \left[\begin{array}{c}
x_{1}-x_{2}\\
0
\end{array}\right]\right\Vert _{2}}{\left\Vert \left[\begin{array}{c}
x_{1}-x_{2}\\
0
\end{array}\right]\right\Vert _{2}}=1,\qquad\forall|x_{1}|,|x_{2}|\in[\delta,1).
\]
When $|x_{1}|\in[0,\delta]$ and $|x_{2}|\in[\delta,1)$, we have
that
\[
y_{\M_{\circ}}(x_{1})-y_{\M_{\circ}}(x_{2})=c\left(1-\frac{|x_{1}|}{\delta}\right),
\]
from which it follows that
\begin{align*}
1\le\frac{\left\Vert \left[\begin{array}{c}
x_{1}\\
y_{\M_{\circ}}(x_{1})
\end{array}\right]-\left[\begin{array}{c}
x_{2}\\
y_{\M_{\circ}}(x_{2})
\end{array}\right]\right\Vert _{2}}{\left\Vert \left[\begin{array}{c}
x_{1}\\
y_{\M}(x_{1})
\end{array}\right]-\left[\begin{array}{c}
x_{2}\\
y_{\M}(x_{2})
\end{array}\right]\right\Vert _{2}}=\frac{\left\Vert \left[\begin{array}{c}
x_{1}-x_{2}\\
c\left(1-|x_{1}|/\delta\right)
\end{array}\right]\right\Vert _{2}}{\left\Vert \left[\begin{array}{c}
x_{1}-x_{2}\\
0
\end{array}\right]\right\Vert _{2}} & \le\sqrt{1+\frac{c^{2}}{\delta^{2}}}
\end{align*}
for all $|x_{1}|\in[0,\delta]$ and $|x_{2}|\in[\delta,1)$. (Above, the ratio is maximized when $x_2=\delta \cdot \text{sign}(x_1)$.)
Finally, when both points are near the origin, we may again confirm that
\[
1\le\frac{\left\Vert \left[\begin{array}{c}
x_{1}\\
y_{\M_{\circ}}(x_{1})
\end{array}\right]-\left[\begin{array}{c}
x_{2}\\
y_{\M_{\circ}}(x_{2})
\end{array}\right]\right\Vert _{2}}{\left\Vert \left[\begin{array}{c}
x_{1}\\
y_{\M}(x_{1})
\end{array}\right]-\left[\begin{array}{c}
x_{2}\\
y_{\M}(x_{2})
\end{array}\right]\right\Vert _{2}}
=
\frac{\left\|
\left[
\begin{array}{c}
x_1-x_2\\
-\frac{c}{\delta}\cdot \left( |x_1|-
|x_2| \right)
\end{array}
\right]
\right\|_2}
{\left\|
\left[
\begin{array}{c}
x_1-x_2\\
0
\end{array}
\right]
\right\|_2}
\le\sqrt{1+\frac{c^{2}}{\delta^{2}}}
\]
for all $|x_{1}|,|x_{2}|\le\delta$. (Above, we again invoked the symmetry of $\M_\circ$ about the vertical axis.) Overall, since $\sqrt{1+\frac{c^{2}}{\delta^{2}}} \le 1 + \frac{c}{\delta}$,  we arrive at
\[
\left|\frac{\left\Vert \left[\begin{array}{c}
x_{1}\\
y_{\M_{\circ}}(x_{1})
\end{array}\right]-\left[\begin{array}{c}
x_{2}\\
y_{\M_{\circ}}(x_{2})
\end{array}\right]\right\Vert _{2}}{\left\Vert \left[\begin{array}{c}
x_{1}\\
y_{\M}(x_{1})
\end{array}\right]-\left[\begin{array}{c}
x_{2}\\
y_{\M}(x_{2})
\end{array}\right]\right\Vert _{2}}-1\right|\le\frac{c}{\delta},\qquad\forall x_{1},x_{2}\in(-1,1).
\]
Owing to (\ref{eq:M* n Mo}), we conclude that
\begin{equation}
\left|\frac{\left\Vert \left[\begin{array}{c}
x_{1}\\
y_{\M_{*}}(x_{1})
\end{array}\right]-\left[\begin{array}{c}
x_{2}\\
y_{\M_{*}}(x_{2})
\end{array}\right]\right\Vert _{2}}{\left\Vert \left[\begin{array}{c}
x_{1}\\
y_{\M}(x_{1})
\end{array}\right]-\left[\begin{array}{c}
x_{2}\\
y_{\M}(x_{2})
\end{array}\right]\right\Vert _{2}}-1\right|\le\frac{c}{\delta},\qquad\forall x_{1},x_{2}\in(-1,1).\label{eq:iso of M*}
\end{equation}

{\bf Completing the Construction.} Putting (\ref{eq:rch of M*}) and (\ref{eq:iso of M*}) together, we deduce the following. There exists a smooth map $\Phi:\R^{2}\rightarrow\R^{2}$ such that $\M_{*}=\Phi(\M)$. Moreover, $\Phi(\cdot)$ is a near-isometry on $\M$ with isometry constants $u=1+\frac{c}{\delta}$ and $l=1$. In particular, since $c<\delta$ by design, $u< 2$ can be made arbitrarily close to one.  Nevertheless,
\[
\mbox{rch}\left(\Phi(\M)\right)=\mbox{rch}\left(\M_{*}\right) \le \frac{\delta}{\sqrt{2}}\ll\infty=\mbox{rch}(\M).
\]
This argument establishes that, in general, it is not possible to lower bound $\mbox{rch}\left(\Phi(\M)\right)$ in terms of $\mbox{rch}(\M)$. Since the mapping $\Phi(\cdot)$ is invertible on $\M$, we note that the entire argument can be reversed, and so in general it is also not possible to upper bound $\mbox{rch}\left(\Phi(\M)\right)$ in terms of $\mbox{rch}(\M)$.

\section{Proof of Theorem \label{sec:Proof-of-Theorem linear Phi}\ref{thm:(Near-Isometry) rch preserved} \\ (Reach Under Linear Near-Isometry)}

For the sake of brevity, we will use the notation $\Phi$ instead of $\Phi_{l,u}$, the matrix representation of the linear map $\Phi_{l,u}(\cdot)$. Let $\Phi=U\Sigma V^{*}\in\R^{m\times n}$ be the singular value decomposition of $\Phi$, with orthobasis $U\in\mathbb{R}^{m\times m}$,   $\Sigma\in\mathbb{R}^{m\times n}$ defined as
\begin{equation}
\Sigma = \left[ \begin{array}{cccc} \sigma_1 & & & \\ & \sigma_2 & & \\ & & \ddots & \\ & & & \sigma_m \end{array} 0_{m\times(n-m)} \right] \in \R^{m\times n}, \label{eq:sigmafull}
\end{equation}
and orthobasis $V \in \mathbb{R}^{n\times n}$. In particular, when $m=n$, $\Sigma$ is a square and diagonal matrix. Note that $\Phi \M = U \Sigma V^{*} \M$. Since $V^{*}$ acts as an isometry on all of $\R^{n}$, it follows from Proposition~\ref{lem:exact iso} that
\begin{equation}
\label{eq:M' to M}
\mbox{rch}(\M'):=\mbox{rch}(V^{*}\M) = \mbox{rch}(\M),
\end{equation}
where we set $\M' := V^{*} \M$ to keep the notation compact.
Thus, it suffices to consider the action of $U \Sigma$ on the rotated manifold $\M'$.
%To keep our notation compact, we set $\M' := V^{*} \M$ for short.
Moreover, since $U$ acts as an isometry on all of $\R^{m}$, we can apply Proposition~\ref{lem:exact iso} again and conclude that
\[
\mbox{rch}(\Phi \M) = \mbox{rch}(U \Sigma V^{*} \M) = \mbox{rch}(U \Sigma \M') = \mbox{rch}(\Sigma \M').
\]
Thus, it suffices to merely consider the action of $\Sigma$ on $\M'$. For any pair of points $x, y \in M'$, note that $Vx, Vy \in \M$, that
\[
\|x - y\|_{2} = \|Vx - Vy\|_{2},
\]
and that
\[
\|\Sigma x - \Sigma y\|_{2} = \|U \Sigma V^{*} V x - U \Sigma V^{*} V y\|_{2} =  \|\Phi V x - \Phi V y\|_{2}.
\]
Therefore, it follows from~\eqref{eq:near iso} that
\begin{equation}
l\cdot\|x-y\|_{2}\le\|\Sigma x - \Sigma y\|_{2}\le u\cdot\|x-y\|_{2},\qquad\forall x,y\in\M'.\label{eq:nearisomprime}
\end{equation}
To summarize,~\eqref{eq:nearisomprime} states that $\Sigma$ acts as a near-isometry on $\M'$, and we wish to bound $\mbox{rch}(\Sigma \M')$ in terms of $\mbox{rch}(\M')$.

Using~\eqref{eq:sigmafull}, we can factor $\Sigma$ as follows:
\begin{equation}
\Sigma = \Sigma' \left[\begin{array}{cc} I_{m} & 0_{m\times(n-m)}\end{array}\right] =: \Sigma' \cdot I_{m,n} \in \R^{m\times n}, \label{eq:def of Imn}
\end{equation}
where $\Sigma' \in \R^{m\times m}$ is a square diagonal matrix containing the nonzero singular values of $\Phi$, and $I_{m,n}\in\R^{m\times n}$ is a rectangular matrix comprised of the $m \times m$ identity and an $m \times (n-m)$ matrix of zeros. In particular, when $m=n$, then $I_{m,n}=I_n$ is the identity matrix (and hence an exact isometry). By assumption, the diagonal entries of $\Sigma'$ belong to the interval $[\sigma_{\min},\sigma_{\max}]$, so that
\begin{equation}
\label{eq:iso on Sigma}
\sigma_{\min}\|g\|_2 \le \|\Sigma' g\|_2 \le \sigma_{\max}\|g\|_2,\qquad \forall g\in\R^m.
\end{equation}
In words, $\Sigma'$ is a bi-Lipschitz map on $\R^m$. If $m<n$, the fact that $\Sigma$ and $\Sigma'$ are bi-Lipschitz maps on $\M'$ and $\R^m$, respectively, implies that $I_{m,n}\in\R^{m\times n}$ is a bi-Lipschitz map on $\M'$. Indeed, it follows from (\ref{eq:nearisomprime}) and (\ref{eq:iso on Sigma})
that
\[
\sigma_{\max}\left\Vert I_{m,n}(x-y)\right\Vert _{2}\ge\left\Vert \Sigma'\cdot I_{m,n}(x-y)\right\Vert _{2}=\left\Vert \Sigma(x-y)\right\Vert _{2}\ge l\|x-y\|_{2},\qquad\forall x,y\in\M',
\]
and consequently,
\begin{equation}
\left\Vert I_{m,n}(x-y)\right\Vert _{2}\ge\frac{l}{\sigma_{\max}}\|x-y\|_{2},\qquad\forall x,y\in\M'.
\label{eq:iso on Imn}
\end{equation}
We now proceed in two steps to prove the claim in Theorem \ref{thm:(Near-Isometry) rch preserved}.

\subsection{First Step}
\label{sec:bigprooffirst}

The first step is to compare $\mbox{rch}(I_{m,n}\M')$ with $\mbox{rch}(\M')$. When $m=n$,  $I_{m,n}=I_n$ is the identity matrix and $I_{m,n}\M'=\M'$. Consequently, $\mbox{rch}(I_{m,n}\M')=\mbox{rch}(\M')$.

We therefore focus on the case where $m<n$. Let $\pi_{m}$ denote the $m$-dimensional subspace spanning the first $m$ coordinates in $\R^{n}$, and let $P_{m}x\in\R^{n}$ be the projection of $x$ onto the subspace $\pi_{m}$. Note that $P_{m}\M'$ (the projection of $\M'$ onto the subspace $\pi_{m}$) is isometric to $I_{m,n}\M'$. By Proposition~\ref{lem:exact iso}, we conclude that $\mbox{rch}(P_{m}\M')=\mbox{rch}(I_{m,n}\M')$. Therefore, we shift our attention to calculating $\mbox{rch}(P_{m}\M')$.

\begin{figure}[t]
\centering
\includegraphics[scale=0.4]{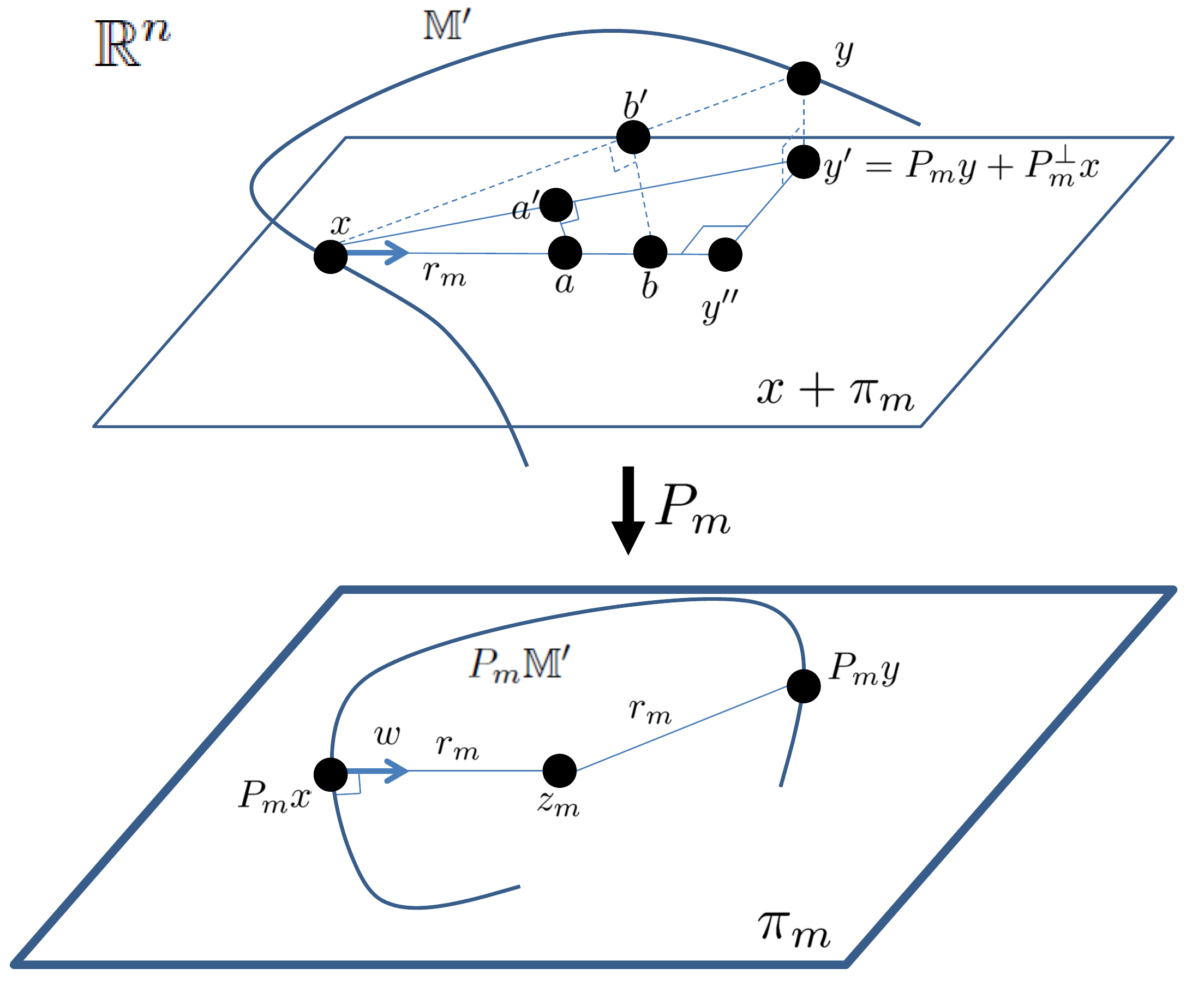}
\caption{\small\sl Proof construction in Appendix \ref{sec:bigprooffirst}.}
\label{fig:bigproof}
\end{figure}

For an arbitrary point $x\in\M'$, consider $P_{m}x$ and an arbitrary unit-length normal vector $w\in\mathbb{N}_{P_{m}x}P_{m}\M'\cap\pi_{m}$, i.e., $w$ is a normal vector at $P_{m}x$ but along $\pi_{m}$.  (The restriction of $w$ to the subspace $\pi_m$  is due to our interest in computing $\mbox{rch}(P_m\M') \subset \pi_m$.)
The constructions mentioned throughout this proof are illustrated in Figure~\ref{fig:bigproof}. For $r_{m}>0$ to be set later, consider the point $z_{m}=P_{m}x+r_{m}\cdot w\in\pi_{m}$. For small enough $r_{m}$, we wish to show that $P_{m}x$ is the unique nearest point to $z_{m}$ on $P_{m}\M'$. To that end, if it at all exists, we consider another point $y\in\M'$ such that
\begin{equation}
r_{m}=\left\Vert z_{m}-P_{m}x\right\Vert _{2}=\left\Vert z_{m}-P_{m}y\right\Vert _{2}.\label{eq:equi dist}
\end{equation}
Because $w=P_{m}w$ is supported only on the first $m$ coordinates, $w$ is also normal to $\M'$, i.e.,
\[
w\in\mathbb{N}_{P_{m}x}P_{m}\M'\Longrightarrow w\in\N_{x}\M'.
\]
Consider the points
\[
a:=x+r_{m}\cdot w,\qquad b:=x+\mbox{rch}(\M')\cdot w,
\]
and note that $a,b\in x+\pi_{m}$. (Here, $x+\pi_{m}$ is the affine subspace parallel to $\pi_{m}$ and passing through $x$.) We keep in mind that, by Definition \ref{def:rch}, $\|b-y\|_{2}>\mbox{rch}(\M')=\|b-x\|_{2}$.

Additionally, let
\[
y':=P_{m}y+P_{m}^{\perp}x\in x+\pi_{m},
\]
be the projection of $y$ onto the affine plane $x+\pi_{m}$. Let also $y''$ be the projection of $y'$ onto the line $\overline{xa}$. Note that $y'$ is the nearest point on the subspace $x+\pi_{m}$ to $y$ and, in turn, $y''$ is the nearest point on the line $\overline{xa}\subset x+\pi_{m}$ to $y'$. We conclude that $y''$ is the nearest point to $y$ on the line $\overline{xa}$. In particular,
\begin{equation}
\overline{yy'}\perp(x+\pi_{m}),\qquad\overline{y'y''}\perp\overline{xa},\qquad\overline{yy''}\perp\overline{xa}.\label{eq:orths}
\end{equation}
Also, let $a'\in x+\pi_{m}$ be the projection of $a$ onto the line $\overline{xy'}$. Finally, let $b'\in \overline{xy}$ be the projection of $b$ onto the line $\overline{xy}$.
From (\ref{eq:equi dist}), it follows that $\|x-a'\|_{2}=\|y'-a'\|_{2}$, and that
\begin{equation}
r_{m} = \frac{\left\Vert x-a'\right\Vert _{2}}{\sin\left(\widehat{a'ax}\right)} =\frac{\frac{1}{2}\left\Vert x-y'\right\Vert _{2}}{\sin\left(\widehat{a'ax}\right)} =\frac{\left\Vert P_{m}(x-y)\right\Vert _{2}}{2\sin\left(\widehat{a'ax}\right)} =\frac{\left\Vert I_{m,n}(x-y)\right\Vert _{2}}{2\sin\left(\widehat{a'ax}\right)} \ge\frac{\frac{l}{\sigma_{\max}}\|x-y\|_{2}}{2\sin\left(\widehat{a'ax}\right)},\label{eq:mid -1}
\end{equation}
where $\widehat{a'ax}$ is, of course, the angle formed by the lines $\overline{a'a}$ and $\overline{xa}$ (facing $\overline{xy'}$), and where the inequality in~\eqref{eq:mid -1} follows from~\eqref{eq:iso on Imn}. In order to control $r_{m}$, we next calculate $\sin(\widehat{a'ax})$.

The geometry of the problem forces that
\begin{equation}
\widehat{a'ax}=\widehat{xy'y''},\label{eq:mid 0.5}
\end{equation}
and that
\begin{equation}
\overline{y'y''}\perp\overline{xa}\Longrightarrow\cos\left(\widehat{xy'y''}\right)=\frac{\left\Vert y'-y''\right\Vert _{2}}{\left\Vert y'-x\right\Vert _{2}},\label{eq:mid 0}
\end{equation}
by (\ref{eq:orths}). Therefore, instead of calculating $\sin(\widehat{a'ax})$, our plan is to find $\cos(\widehat{xy'y''})$ first. To that end, note that
\begin{equation}
\overline{yy'}\perp(x+\pi_{m})\Longrightarrow\left\Vert y'-y''\right\Vert _{2}^{2}=\left\Vert y''-y\right\Vert _{2}^{2}-\left\Vert y'-y\right\Vert _{2}^{2},\label{eq:mid 1}
\end{equation}
by (\ref{eq:orths}) again. % But $\|y''-y\|_{2}$ can be computed by focusing on the triangle formed by the points $x$, $y$, and $b$.
After recalling that $\|x-b\|_{2}=\mbox{rch}(\M')<\|y-b\|_{2}$, we write that
\begin{align*}
\left\Vert y''-y\right\Vert _{2}^{2} & =\|x-y\|_{2}^{2}\cdot\sin^{2}\left(\widehat{yxa}\right)\qquad\left(\overline{yy''}\perp\overline{xa}\,\mbox{ from (\ref{eq:orths})}\right)\\
& = \|x-y\|_{2}^{2}\cdot\cos^{2}\left(\widehat{xbb'}\right)\qquad
\left(
\widehat{yxa}+\widehat{xbb'}=\frac{\pi}{2}
\right)
\nonumber\\
& = \|x-y\|_{2}^{2}\cdot\left( 1- \sin^{2}\left(\widehat{xbb'}\right) \right)\nonumber\\
& =
\|x-y\|_{2}^{2}\cdot\left( 1- \frac{\left\|x-b' \right\|_2^2}{\mbox{rch}\left(\M' \right)^2} \right)
\qquad
\left( \left\| x-b\right\|_2= \mbox{rch}(\M')\right)
\nonumber\\
 & \ge \|x-y\|_{2}^{2}\cdot\left[1-\left(\frac{\|x-y\|_{2}}{2\cdot\mbox{rch}(\M')}\right)^{2}\right].
 \qquad\left(\|x-b\|_{2}\le\|y-b\|_{2}
\Rightarrow
\|x-b'\|_{2}\le\|y-b'\|_{2}
 \right)
\end{align*}
Substituting the expression for $\left\Vert y''-y\right\Vert _{2}$ above back in (\ref{eq:mid 1}), we arrive at
\begin{equation*}
\left\Vert y'-y''\right\Vert _{2}^{2} \ge\|x-y\|_{2}^{2}\cdot\left[1-\left(\frac{\|x-y\|_{2}}{2\cdot\mbox{rch}(\M')}\right)^{2}\right]-\left\Vert y'-y\right\Vert _{2}^{2}.
\end{equation*}
In turn, plugging the expression for $\left\Vert y'-y''\right\Vert _{2}$ above back into (\ref{eq:mid 0}) yields
\[
\cos^{2}\left(\widehat{xy'y''}\right)\ge\frac{\|x-y\|_{2}^{2}\cdot\left[1-\left(\frac{\|x-y\|_{2}}{2\cdot\mbox{rch}(\M')}\right)^{2}\right]-\left\Vert y'-y\right\Vert _{2}^{2}}{\left\Vert y'-x\right\Vert _{2}^{2}}.
\]
By (\ref{eq:mid 0.5}), then, we obtain that
\begin{align*}
\sin^{2}\left(\widehat{a'ax}\right) & =\sin^{2}\left(\widehat{xy'y''}\right)\\
 & =1-\cos^{2}\left(\widehat{xy'y''}\right)\\
 & \le1-\frac{\|x-y\|_{2}^{2}\cdot\left[1-\left(\frac{\|x-y\|_{2}}{2\cdot\mbox{rch}(\M')}\right)^{2}\right]-\left\Vert y'-y\right\Vert _{2}^{2}}{\left\Vert y'-x\right\Vert _{2}^{2}}\\
 & =\frac{\left\Vert y'-x\right\Vert _{2}^{2}+\left\Vert y'-y\right\Vert _{2}^{2}-\|x-y\|_{2}^{2}\cdot\left[1-\left(\frac{\|x-y\|_{2}}{2\cdot\mbox{rch}(\M')}\right)^{2}\right]}{\left\Vert y'-x\right\Vert _{2}^{2}} \\
 & =\frac{\|x-y\|_{2}^{2}-\|x-y\|_{2}^{2}\cdot\left[1-\left(\frac{\|x-y\|_{2}}{2\cdot\mbox{rch}(\M')}\right)^{2}\right]}{\left\Vert y'-x\right\Vert _{2}^{2}}
\qquad\left(\overline{yy'}\perp(x+\pi_{m})\right)
 \\
 & =\frac{\|x-y\|_{2}^{2}\cdot\left(\frac{\|x-y\|_{2}}{2\cdot\mbox{rch}(\M')}\right)^{2}}{\left\Vert y'-x\right\Vert _{2}^{2}}.
\end{align*}
Plugging the above expression for $\sin(\widehat{a'ax})$ back into (\ref{eq:mid -1}), we obtain that
\begin{align*}
r_{m} & \ge\frac{\frac{l}{\sigma_{\max}}\|x-y\|_{2}}{2\sin\left(\widehat{a'ax}\right)}\\
 & \ge\mbox{rch}(\M')\cdot\frac{l}{\sigma_{\max}}\cdot\frac{\left\Vert y'-x\right\Vert _{2}}{\|x-y\|_{2}}\\
 & =\mbox{rch}(\M')\cdot\left(\frac{l}{\sigma_{\max}}\right)^{2},\qquad\mbox{(see (\ref{eq:iso on Imn}))}
\end{align*}
where the last line uses a similar argument to that used in (\ref{eq:mid -1}). Since our choices of $x$ and $w$ were otherwise arbitrary, we conclude that
\begin{equation}
\mbox{rch}\left(I_{m,n}\M'\right)=\mbox{rch}\left(P_{m}\M'\right)\ge\left(\frac{l}{\sigma_{\max}}\right)^{2}\cdot\mbox{rch}(\M').\label{eq:step 1}
\end{equation}

\subsection{Second Step}

So far, we have computed $\mbox{rch}(I_{m,n}\M')$. In this section, we complete the argument by computing $\mbox{rch}(\Sigma'\cdot I_{m,n}\M')=\mbox{rch}(\Sigma\M') = \mbox{rch}(\Phi\M)$. We do so by studying the linear map $\Sigma'$.

As indicated in (\ref{eq:iso on Sigma}), $\Sigma'$ is a bi-Lipschitz map on $\R^{m}$ with constants $\sigma_{\min}\le \sigma_{\max}$. Therefore, intuitively, it should be clear that $\Sigma'$ does not substantially distort the geometry of $I_{m,n}\M'\subset\R^{m}$. Next, we make this notion concrete.

Consider an arbitrary point $I_{m,n}x\in I_{m,n}\M'$. By Definition \ref{def:rch}, one can place an $m$-dimensional (Euclidean) ball of radius $\mbox{rch}(I_{m,n}\M')$ tangent to $I_{m,n}\M'$ at $I_{m,n}x$ that never intersects $I_{m,n}\M'$.
%
% (Though this ball might be tangent to $I_{m,n}\M'$ at points other than $I_{m,n}x$.)
%
Under $\Sigma'$, this ball turns into an ellipsoid. The longest principal axis of this ellipsoid does not exceed $\sigma_{\max}\cdot\mbox{rch}(I_{m,n}\M')$ and its smallest principal axis is no less than $\sigma_{\min}\cdot\mbox{rch}(I_{m,n}\M')$. The reach of this ellipsoid is never less than $\frac{\sigma_{\min}^{2}}{\sigma_{\max}}\cdot\mbox{rch}(I_{m,n}\M')$ as guaranteed by Proposition~\ref{lem:ellipsoidreach}. Therefore, this ellipsoid itself contains an $m$-dimensional ball of radius $\frac{\sigma_{\min}^{2}}{\sigma_{\max}}\cdot\mbox{rch}(I_{m,n}\M')$ that is tangent at $\Sigma' \cdot I_{m,n}x=\Sigma x$ and never intersects the ellipsoid (and hence the manifold $\Sigma\M'$).

It is possible to choose the center of the original Euclidean ball in any direction normal to $I_{m,n}\M'$ at $I_{m,n}x$; the resulting Euclidean balls (which are tangent to $\Sigma x$) trace out all normal directions at $x$. Since the choice of $x$ was otherwise arbitrary, it follows that
\begin{equation}
\mbox{rch}(\Phi\M) = \mbox{rch}(\Sigma\M')=\mbox{rch}\left(\Sigma'\cdot I_{m,n}\M'\right)  \ge\frac{\sigma_{\min}^{2}}{\sigma_{\max}}\cdot\mbox{rch}(I_{m,n}\M').
\label{eq:pre final}
\end{equation}
When $m=n$, in particular, we continue by noting that
\begin{align*}
\mbox{rch}(\Phi\M)  & \ge\frac{\sigma_{\min}^{2}}{\sigma_{\max}}\cdot\mbox{rch}\left(I_{m,n}\M'\right)
\qquad \mbox{(see \eqref{eq:pre final})}
\\
& = \frac{\sigma_{\min}^{2}}{\sigma_{\max}}\cdot\mbox{rch}(\M') \qquad \left(I_{m,n}=I_n \mbox{ is the identity matrix}\right)\\
& = \frac{\sigma_{\min}^{2}}{\sigma_{\max}}\cdot\mbox{rch}(\M)\qquad \mbox{(see \eqref{eq:M' to M})}.
\end{align*}
On the other hand, when $m<n$, we write that
\begin{align*}
\mbox{rch}(\Phi\M)  & \ge\frac{\sigma_{\min}^{2}}{\sigma_{\max}}\cdot\mbox{rch}\left(I_{m,n}\M'\right)
\qquad \mbox{(see \eqref{eq:pre final})}
\\
& \ge\frac{\sigma_{\min}^{2}}{\sigma_{\max}}\cdot\left(\frac{l}{\sigma_{\max}}\right)^{2}\cdot\mbox{rch}(\M') \qquad\mbox{(see (\ref{eq:step 1}))}
 \\
 & = \frac{\sigma_{\min}^{2}}{\sigma_{\max}}\cdot\left(\frac{l}{\sigma_{\max}}\right)^{2}\cdot\mbox{rch}(\M).
\end{align*}
This completes the proof of Theorem \ref{thm:(Near-Isometry) rch preserved}.

%\begin{acknowledgements}
%\end{acknowledgements}


\end{document}